\documentclass[fleqn,usenatbib]{mnras}

\usepackage{newtxtext,newtxmath}
\usepackage{pdflscape} % Landscape pages
\usepackage[T1]{fontenc}
\DeclareRobustCommand{\VAN}[3]{#2}
\let\VANthebibliography\thebibliography
\def\thebibliography{\DeclareRobustCommand{\VAN}[3]{##3}\VANthebibliography}

\def\aj{{AJ}}% 
\def\araa{{ARA\&A}}% 
\def\apj{{ApJ}}% 
\def\apjl{{ApJ}}% 
\def\apjs{{ApJS}}% 
\def\pasp{{PASP}}% 
\def\mnras{{MNRAS}}% 
\newcommand{\kms}{\,km\,s$^{-1}$}
\newcommand{\mstellar}{\ensuremath{M_{\mathrm{stellar}}}}
\newcommand{\omatter}{\ensuremath{\Omega_{\mathrm{M}}}}

\newcommand{\Siii}{\ensuremath{\mathrm{Si}\,\textsc{ii}\,\lambda6355}}
\newcommand{\vsiii}{\ensuremath{v_{\mathrm{Si}\,\textsc{ii}}}}
\newcommand{\Cii}{\ensuremath{\mathrm{C}\,\textsc{ii}\,\lambda6580}}
\newcommand{\Feii}{\ensuremath{[\mathrm{Fe}\,\textsc{ii}}]}

\usepackage[transparent]{graphicx}	% Including figure files
\usepackage{amsmath}	% Advanced maths commands
\usepackage[group-separator={,}]{siunitx}
\title[Local host properties of HV SNe~Ia]{A closer look at the host-galaxy environment of high-velocity Type Ia supernovae}

 \author[Lin, H.-T. et al.]{
 Han-Tang~Lin$^{1}$\footnote[2],
 Yen-Chen~Pan$^{1}$\thanks{E-mail: ycpan@astro.ncu.edu.tw}\footnote[2],
 and Abdurro'uf$^{2,3}$
 \\
 $^{1}$Graduate Institute of Astronomy, National Central University, 300 Jhongda Road, 32001 Jhongli, Taiwan\\
 $^{2}$Center for Astrophysical Sciences, Department of Physics and Astronomy, The Johns Hopkins University, 3400 N Charles St., Baltimore, MD 21218, USA\\
 $^{3}$Space Telescope Science Institute (STScI), 3700 San Martin Drive, Baltimore, MD 21218, USA\\
 }

\pubyear{2023}

\begin{document}
\label{firstpage}
\pagerange{\pageref{firstpage}--\pageref{lastpage}}
\maketitle

% Abstract of the paper
% less than 200 words 
\begin{abstract}
Recent studies suggested that the ejecta velocity of Type~Ia supernova (SN~Ia) is a promising indicator in distinguishing the progenitor systems and explosion mechanisms. By classifying the SNe~Ia based on their ejecta velocities, studies found SNe~Ia with high \Siii\ velocities (HV SNe~Ia; $ v \ga 12000$\,\kms) tend to be physically different from their normal-velocity counterparts (NV SNe~Ia). In this work, we revisit the low-$z$ sample studied in previous work and closely look into the spatially resolved environment local to the site of SN explosion. Our results reveal a possible trend (at $2.4\sigma$ significance) that HV SNe~Ia are likely associated with older stellar populations than NV SNe~Ia. While the trend is inconclusive, the local host-galaxy sample studied in this work is likely skewed toward massive galaxies, limiting the parameter space that we would like to investigate from the original parent sample. Nevertheless, our results do not rule out the possibility that parameters other than the host-galaxy age (such as metallicity) could be the underlying factors driving the differences between HV and NV SNe~Ia due to the limitation of our dataset.
\end{abstract}

% Select between one and six entries from the list of approved keywords.
% Don't make up new ones.

\footnotetext[2]{Co-first authors}

%%%%%%%%%%%%%%%%%%%%%%%%%%%%%%%%%%%%%%%%%%%%%%%%%%

%%%%%%%%%%%%%%%%% BODY OF PAPER %%%%%%%%%%%%%%%%%%
\begin{figure*}
 \centering
  \includegraphics[scale=0.6]{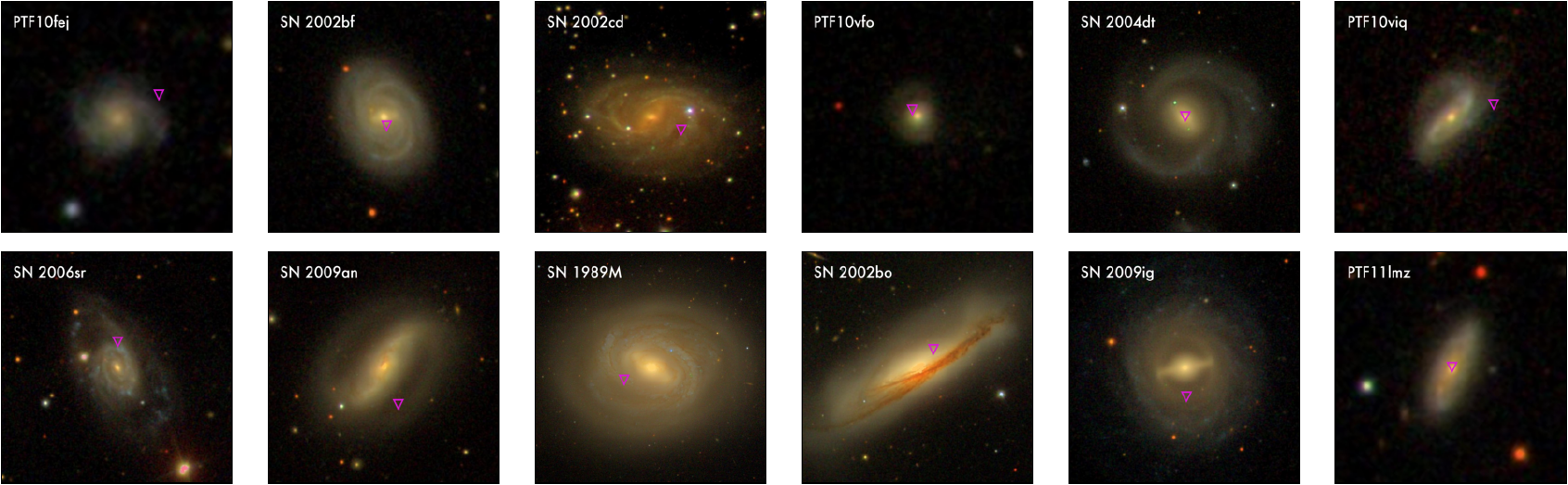}
  \caption{Examples of our SN~Ia host galaxies for the local analyses. The triangle in each panel shows the SN location. The images are generated from the SDSS REST Web Services.}
  \label{fig:marked_samples}
\end{figure*}

\section{Introduction}
\label{sec:introduction}

Type Ia supernovae (SNe~Ia) are attributed to the thermonuclear explosion of a white dwarf (WD) star composed mainly of carbon and oxygen in a close binary system. However, the precise identity of the companion star that donates material to the WD remains an open question. Potential scenarios include the single degenerate model \citep{1973ApJ...186.1007W} and the double degenerate model \citep{1984ApJS...54..335I,1984ApJ...277..355W}, along with several variations based on these two fundamental pictures \citep[e.g.,][]{2018PhR...736....1L,2019NewAR..8701535S,2023RAA....23h2001L}. If normal SNe~Ia originate from multiple pathways, distinguishing between these channels will be critical for their utility in cosmological studies.

Previous studies proposed that the ejecta velocity of SN~Ia, especially the \Siii\ velocity, is a promising indicator in characterizing the observed diversities \citep{2008ApJ...675..626W} and even different progenitor scenarios. This distinction was initially proposed by \citet{2009ApJ...699L.139W}, who noted that the SNe~Ia with high \Siii\ velocity (HV SNe~Ia; $ \vsiii\ \ga 12000$\,\kms) tend to appear redder and exhibit a lower extinction ratio ($R_{V}$) than their normal \Siii\ velocity counterparts (NV SNe~Ia; $ \vsiii\ \la 12000$\,\kms), although some HV~SNe Ia may be intrinsically redder rather than following a different reddening law relative to their normal counterparts \citep{2011ApJ...729...55F}. More recently, \citet{2019ApJ...882..120W} found evidence that HV SNe~Ia tend to display a blue excess in their late-time light curves. They also found variable Na\,\textsc{i} absorption lines in the HV SN~Ia spectra. These observations were attributed to circumstellar dust surrounding the SNe. By studying the polarization of the \Siii\ line, \citet{2019MNRAS.490..578C} found a significant relationship between the polarization degree and \Siii\ velocity, in the sense that SNe with higher ejecta velocities tend to show higher degrees of polarization. This implies the ejecta velocity could be critical in tracing different explosion mechanisms (such as their geometrical structures). Moreover, \citet{2021ApJ...906...99L} showed that HV SNe~Ia are likely to present either very weak or no \Cii\ feature in their early-time spectra, while this feature is always visible for those of NV SNe~Ia. They also unveiled distinctions in the late-time spectroscopic properties of NV and HV SNe~Ia, with all HV SNe in their sample displaying redshifted \Feii\ lines, whereas NV SNe exhibited both blueshifted and redshifted \Feii\ lines. They proposed that the scenarios that include a Helium detonation near the surface of WD could explain the unique properties of HV SNe~Ia (and possibly part of the NV SNe~Ia). By compiling a large spectroscopic sample of up to $z=0.6$, \citet{2022arXiv221106895P} found HV and NV SNe~Ia could have different redshift distributions, with the fraction of HV SNe~Ia likely to be lower at the high-$z$ Universe. They suspected that the progenitor scenarios of HV SNe~Ia may favor a longer delay time than that of NV SNe~Ia.

The study of host galaxies has also been successful in constraining SN physics. In particular, many previous studies have seen significant relations between SN Ia ejecta velocity and host-galaxy properties. \citet{2013Sci...340..170W} found that HV SNe~Ia tend to cluster in the inner regions of their host galaxies. In contrast, NV events exhibit a broader distribution across radial distances. They proposed that HV SNe~Ia are likely associated with metal-rich populations. \citet{2015MNRAS.446..354P}, \citet{2020ApJ...895L...5P}, and \citet{2021ApJ...923..267D} further supported this idea, finding significant evidence that HV SNe~Ia tend to be discovered in more massive galaxies, while their NV counterparts can exist in both low-mass and massive galaxies. They suggested the progenitor systems of HV SNe~Ia could be old and/or metal-rich. Conversely, \citet{2023arXiv230410601N} did not see a significant trend between \Siii\ velocity and global host-galaxy properties, such as stellar mass (\mstellar), age, and star formation rate. However, they saw more significant relations with SN location and Na\,\textsc{i}\,D equivalent widths, with HV SNe~Ia having stronger Na\,\textsc{i}\,D absorption and concentrating toward the center of host galaxies. Their results are consistent with findings from \citet{2013Sci...340..170W,2019ApJ...882..120W}.

In this work, we look into the host-galaxy properties local to the SN, with the goal of investigating the potential differences between HV and NV SNe~Ia that previous studies on the global properties of host galaxies did not reveal. The structure of this paper is as follows. In Section~\ref{sec:data-method}, we introduce our SN~Ia sample and the methodology for determining the local properties of host galaxies. The results are presented in Section~\ref{sec:results}. We further discuss our results in Sections~\ref{sec:discussion} and provide our conclusions in Section~\ref{sec:conclusions}. Throughout this paper, we assume $\mathrm{H_0}=70$\,km\,s$^{-1}$\,Mpc$^{-1}$ and a flat universe with $\omatter=0.3$.

\begin{figure}
\centering
  \includegraphics[scale=0.3]{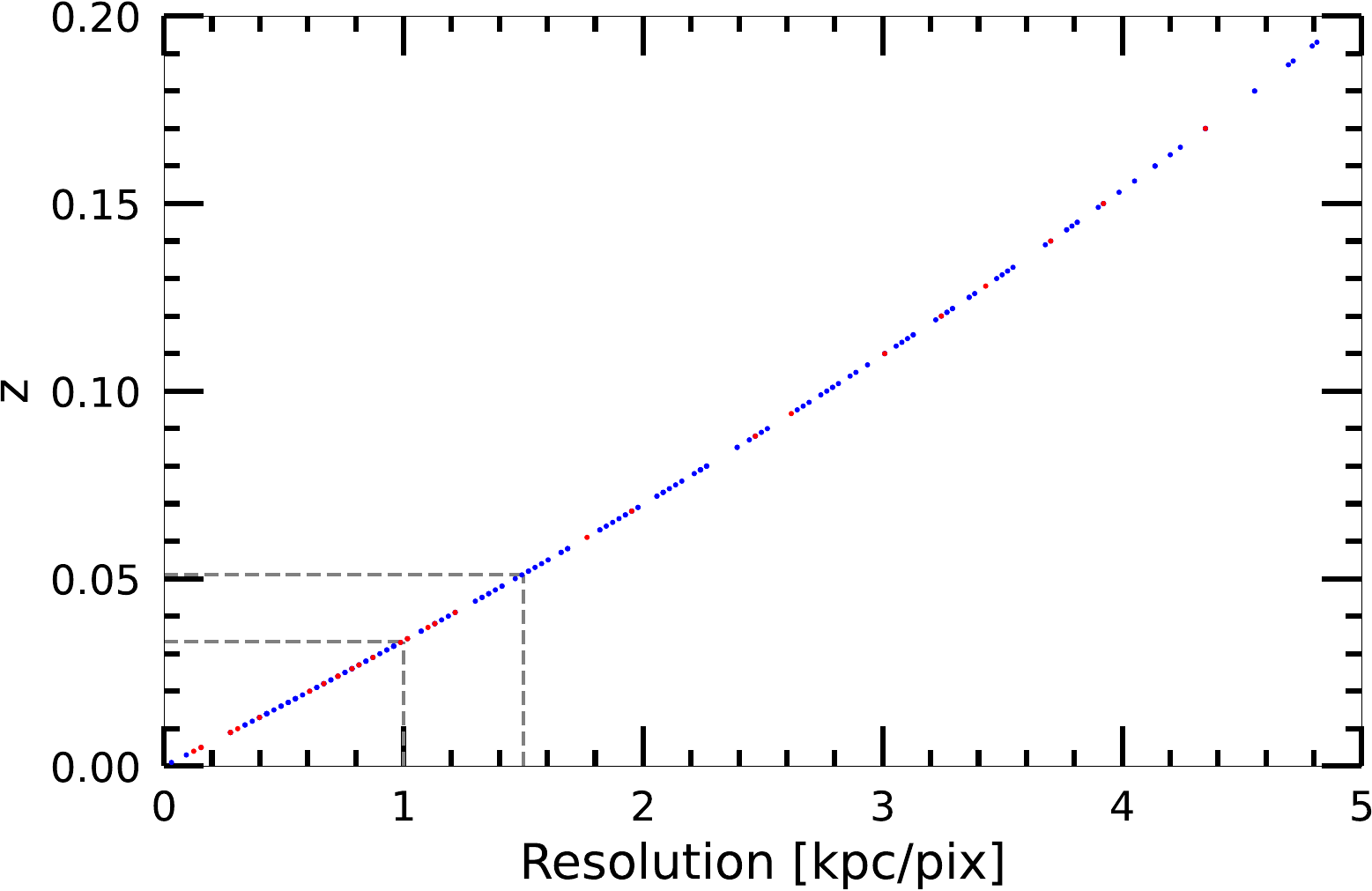}
  \caption{The relationship between redshift and physical scale in a unit of kilo-parsec (kpc) for the stamp images. Given the spatial sampling of our images, the redshift cuts of 0.033 and 0.051 are applied to achieve physical resolutions of 1 and 1.5\,kpc, respectively (as indicated by the dashed lines). The blue and red dots in the figure represent the NV and HV SNe~Ia, respectively.}
  \label{fig:resolution}
\end{figure}

\section{Data and Method}
\label{sec:data-method}

\subsection{SN data}
\label{sec:sn-data}

% Sample Selection 
We use the SN~Ia host-galaxy sample studied in \citet{2020ApJ...895L...5P} as a parent sample in this work. This contains 219 SNe~Ia with both \vsiii\ and global host-galaxy measurements from SDSS imaging data in the local Universe \citep[see the Table~1 in][]{2020ApJ...895L...5P}, primarily from the Palomar Transient Factory \citep[PTF;][]{2009PASP..121.1395L,2009PASP..121.1334R} and Berkeley SN~Ia Program \citep[BSNIP;][]{2012MNRAS.425.1789S}. This nearby sample is ideal for investigating the spatially resolved environment local to the site of SN explosion. The near-peak \Siii\ velocities were already measured from \citet{2020ApJ...895L...5P}. The details of the sample selection and velocity measurement can be found in \citet{2014MNRAS.444.3258M}. Briefly speaking, we obtained the near-peak sample with the optical spectra observed within three days from the peak luminosity. The process begins by initially correcting the SN spectrum to the rest frame. The continuum regions on both sides of the feature are defined through visual inspection, and a pseudo-continuum is then fitted across the absorption feature. Normalization of the feature is achieved by dividing it by the pseudo-continuum. A Gaussian fit is then applied to the normalized \Siii\ line in velocity space, providing the resulting fit with velocity and pseudo-equivalent widths for the \Siii\ feature.

\subsection{Host-galaxy data}
\label{sec:host-data}

\begin{figure*}
\centering
  \includegraphics[scale=0.4]{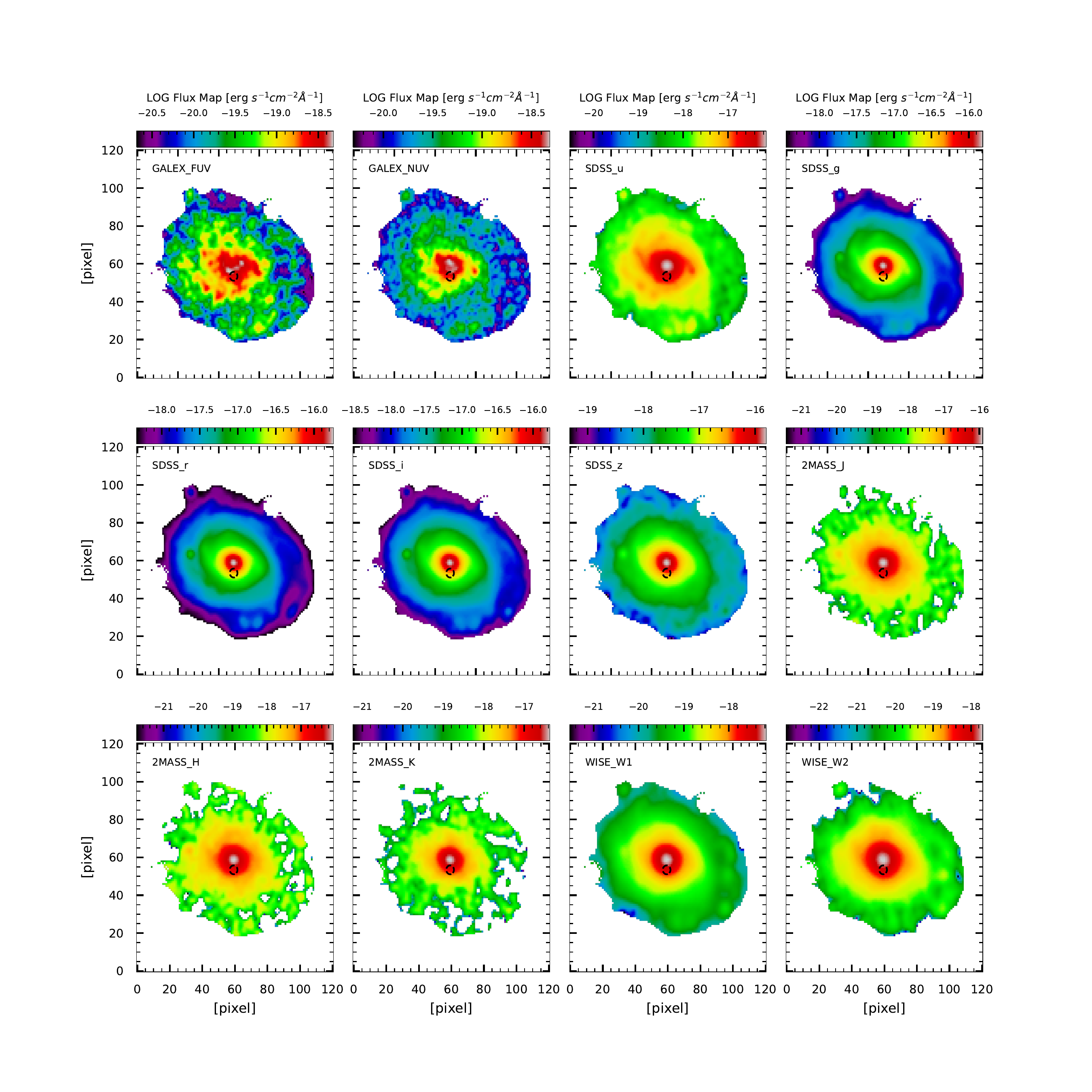}\\
  \vspace{0.4cm}
  \includegraphics[scale=0.55]{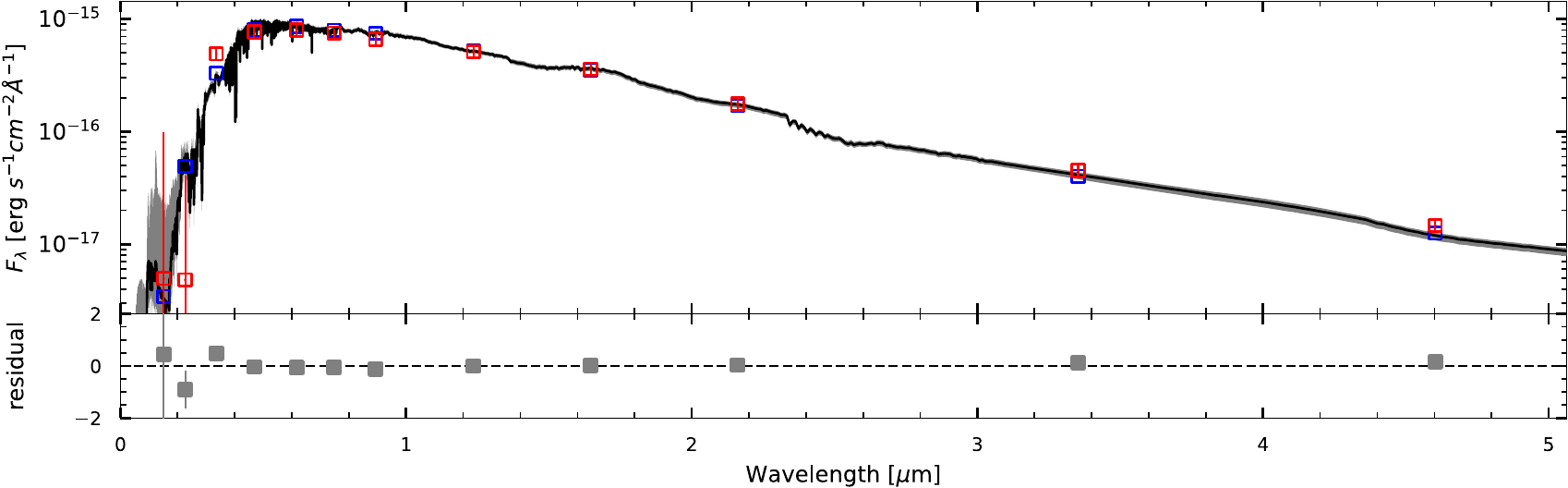}
  \caption{The small panels show the flux maps of the host galaxy of SN~2006or in 12 bands, which are processed from the imaging data of GALEX, SDSS, 2MASS, and WISE. Each panel is color-coded in terms of the flux (in logarithmic scale). A circular aperture at the SN location with a radius of 1.5\,kpc is overplotted in each panel. The bottom panel shows the SED of the galaxy. The observed photometry extracted from the 1.5\,kpc aperture in each filter is shown in red squares. The best-fit photometry (i.e., with the lowest reduced $\chi^2$) and template spectrum from the model are shown in blue squares and black curve, respectively. The grey shaded area represents the uncertainty of the best-fit model.  The residual between observation and model is also shown, which is defined as the difference between observed flux and best-fit flux normalized by the best-fit flux.} 
  \label{fig:seds}
\end{figure*}

To carry out the spectral energy distribution (SED) fitting on spatially-resolved scales of the host galaxies, we collect archival images with a wide range of wavelengths, spanning from ultraviolet (UV) to infrared (IR).

% UV 
For the UV images, we retrieve the data from the Galaxy Evolution Explorer \citep[GALEX;][]{2007ApJS..173..682M}, including both the FUV (1350-1780\,\AA) and NUV (1770-2730\,\AA) bands. We prioritize our pipeline by accessing the data from the Medium Imaging Survey (MIS) over the All-Sky Imaging Survey (AIS) to obtain better spatially resolved photometry over the optical region of the galaxies as the MIS survey is much deeper than the AIS survey. This survey achieves $5\sigma$ limiting magnitudes of 22.6 and 22.7 in FUV and NUV, respectively. These UV images have a pixel resolution of 1.5\arcsec. The full-width at half maximum (FWHM) of the point-spread function (PSF) of FUV and NUV imaging are 4.2\arcsec\ and 5.3\arcsec, respectively.

% Optical
The optical images are obtained with the Sloan Digital Sky Survey (SDSS). The SDSS imaging survey has five filters ($u$, $g$, $r$, $i$, and $z$), which cover wavelengths from 3551 to 8932\,\AA\ \citep{1998AJ....116.3040G,2000AJ....120.1579Y}. The depth of the imaging are: $u=22.0$, $g=22.2$, $r=22.2$, $i=21.3$, and $z=20.5$\,mag \citep{2004AJ....128..502A}. The pixel resolution of the imaging data is 0.396\arcsec, and the median seeing is around 1.3\arcsec\ \citep[e.g.,][]{2011MNRAS.417.1350R}

% NIR 
We access the near-IR (NIR) data from the Two Micron All-Sky Survey \citep[2MASS;][]{2006AJ....131.1163S}. The imaging survey has three filter bands: $J$-band (1.25$\mu$m), $H$-band (1.65$\mu$m), and $Ks$-band (2.15$\mu$m). The pixel resolution of the 2MASS imaging data is 1.0\arcsec. The seeing ranges from $\sim2.5$\arcsec\ to 3\arcsec. The survey achieved point-source sensitivities (at a signal-to-noise ratio of 10) of 15.8, 15.1, and 14.3\,mag for $J$, $H$, and $Ks$, respectively.

% IR 
To obtain a wide wavelength coverage, we add IR images from the Wide-field Infrared Survey Explorer \citep[WISE;][]{2010AJ....140.1868W}, which mapped the whole sky in four bands: 3.4\,$\mu$m (W1), 4.6\,$\mu$m (W2), 12\,$\mu$m (W3), and 22\,$\mu$m (W4). We use the imaging data product from the AllWISE database. We only use W1 and W2 images in our analysis for reasons explained in Section~\ref{sec:fitting}. The imaging in W1 and W2 have PSF FWHM of 6.1\arcsec\ and 6.4\arcsec, respectively. The WISE survey achieved a $5\sigma$ point-source sensitivities better than 0.08 and 0.11\,mJy for W1 and W2 bands, respectively.

Figure~\ref{fig:marked_samples} shows examples of our SN~Ia host galaxies in the composite SDSS images.

\subsection{Image processing and SED fitting of the host galaxies}
\label{sec:fitting}

We use the \textsc{piXedfit} package \citep{2021ApJS..254...15A,2022ascl.soft07033A,2023ApJ...945..117A} for performing image processing and SED fitting to the host galaxies. \textsc{piXedfit} is a Python package designed to perform spatially resolved SED fitting using multi-band imaging data to derive spatially resolved stellar population properties of a galaxy. Detailed descriptions of \textsc{piXedfit} are given in \citet{2021ApJS..254...15A}, including thorough testing of its performance in inferring robust properties of the stellar population. Below we briefly summarize the procedure of image processing and SED fitting of our SN host galaxies.

Before the SED fitting can be performed, we must ensure that all the images have been convolved and resampled into the same spatial resolution, sampling, and projection. This is done with the image processing module in \textsc{piXedfit}. Briefly speaking, the whole image processing includes background subtraction, PSF matching, spatial resampling and reprojection, defining the region of interest, calculating flux and flux uncertainties of pixels, and correcting for the extinction due to the foreground Galactic dust. The science images, variance images (constructed using \textsc{piXedfit}), the coordinates of the host galaxy and SN, redshift, and the desired size for the final images are provided here. The final stamp images with uniform PSF size and spatial sampling are then generated at the end of this process. The stamp images in this work have a PSF FWHM of 6.4\arcsec\ and a pixel size of 1.5\arcsec.

After image processing, we measure aperture photometry at the SN coordinate with \textsc{piXedfit}. Two apertures with radii equivalent of 1 and 1.5\,kpc are used, and total fluxes from pixels within those apertures are calculated. To resolve the SN local environment with our apertures, we limit the redshifts of our sample to be lower than 0.033 and 0.051 for $r=1$\,kpc and 1.5\,kpc\ samples, respectively. Following this redshift constraint, the $r=1$\,kpc sample contains 82 SNe, while the $r=1.5$\,kpc sample contains 107 SNe. To better constrain the host-galaxy properties and disentangle the degeneracy between age and metallicity, we require that the imaging data across all filters (i.e., those indicated in Section~\ref{sec:host-data}) be available at the SN location. We further exclude $\sim30$ SNe due to either unidentified host galaxies or a lack of overlap with the SN location in the imaging data. This gives a final sample of 57 and 71 SNe for $r=1$\,kpc and 1.5\,kpc\ samples, respectively. The relation between the redshift and spatial resolution of our sample can be found in Figure~\ref{fig:resolution}.

The SED fitting is then performed to the aperture photometry measured from the SN region.
\textsc{piXedfit} uses the Flexible Stellar Population Synthesis \citep[FSPS;][]{2009ApJ...699..486C} code through a Python wrapper Python-FSPS \citep{2014zndo.....12157F} in modeling the SED of a stellar population. The modeling incorporates four emission sources, including stellar, nebular, dust, and AGN dusty torus emissions. Since the dust and AGN torus models can only be precisely constrained with FIR data, we turn off both models due to the limited wavelength coverage in this work (i.e., from UV to MIR). Thus, we do not include the WISE W3 and W4 data in our analysis as they are mostly dominated by dust emission. Throughout this work, we adopt the Chabrier initial mass function \citep{2003PASP..115..763C} and MILES stellar library \citep{2006MNRAS.371..703S} when performing the SED fitting. The dust attenuation law described by \citet{2000ApJ...533..682C} is used for dust extinction correction. We assume the delayed-tau star formation history (SFH) model. We have also tried the double power-law SFH model and found that it does not change the results and conclusions in this paper. The SED fitting in \textsc{piXedfit} applies a fully Bayesian inference technique with two posterior sampling methods available: the Markov Chain Monte Carlo (MCMC) and random dense sampling of parameter space (RDSPS). In our analysis, we use the MCMC method as it is expected to yield more accurate and reliable results (although with the expense of computation costs). 

After the SED fitting process, the host-galaxy parameters (such as SFR, mass-weighted stellar age and metallicity) and their uncertainties are obtained from the median, 16th, and 84th percentiles of the posterior probability distributions constructed from the sampler chains (in the case of MCMC fitting method). The host-galaxy parameters measured in this work are listed in Table~\ref{sample_1.5_1} to \ref{sample_1.0_2}. Examples of stamp images (obtained from the image processing) and SED fitting results are shown in Figure~\ref{fig:seds}. We present the flux maps for SN~2006or host galaxy in 12 filter bands spanning from UV to MIR. The SN location is indicated by the black circle. A panel of SED shows the observed aperture photometry at the SN location and the best-fit SED model.

\begin{figure*}
\centering
  \vspace{0.4cm}
  \includegraphics[scale=0.55]{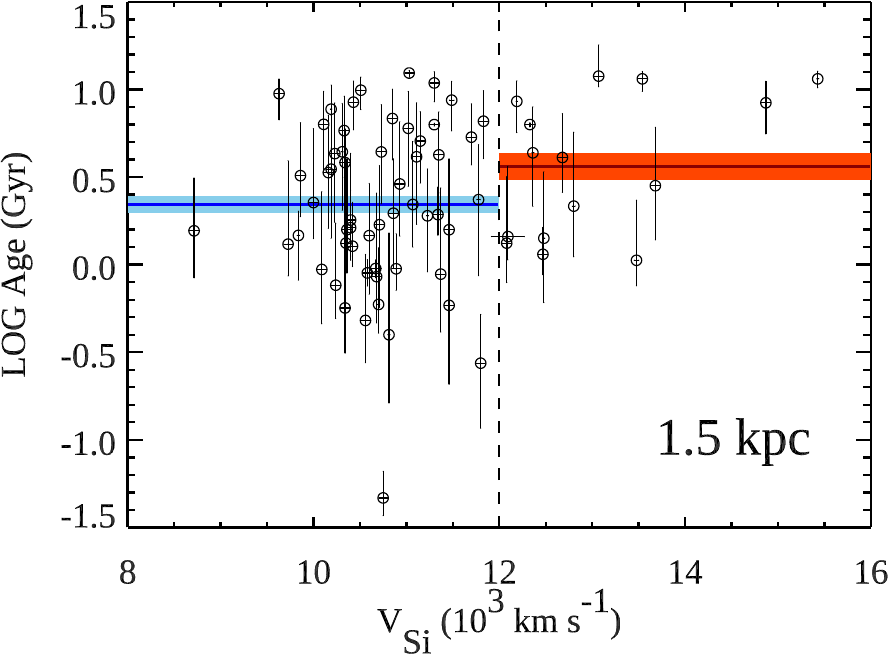}
  \hspace{0.5cm}
  \includegraphics[scale=0.55]{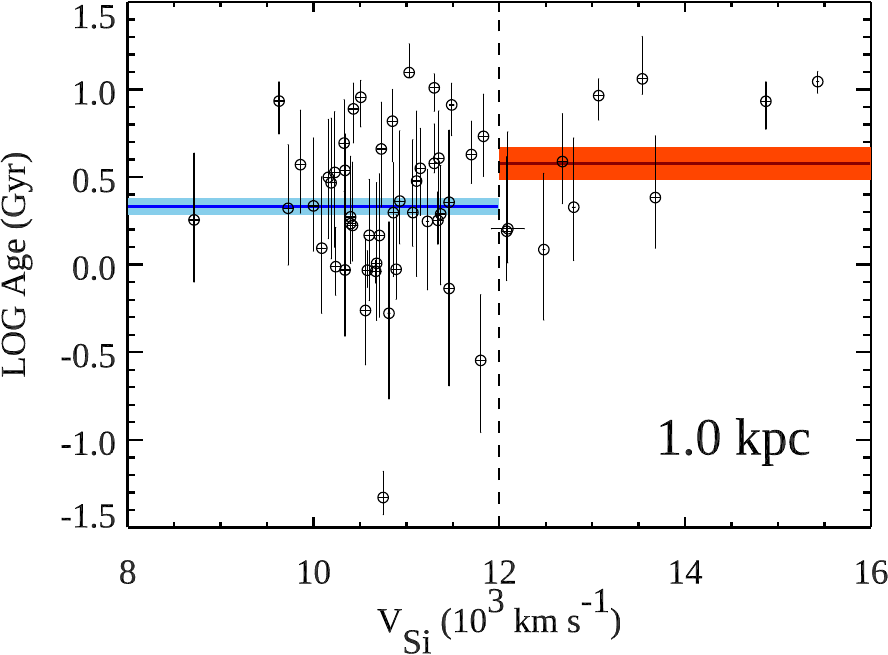}\\
  \vspace{0.6cm}
  \includegraphics[scale=0.55]{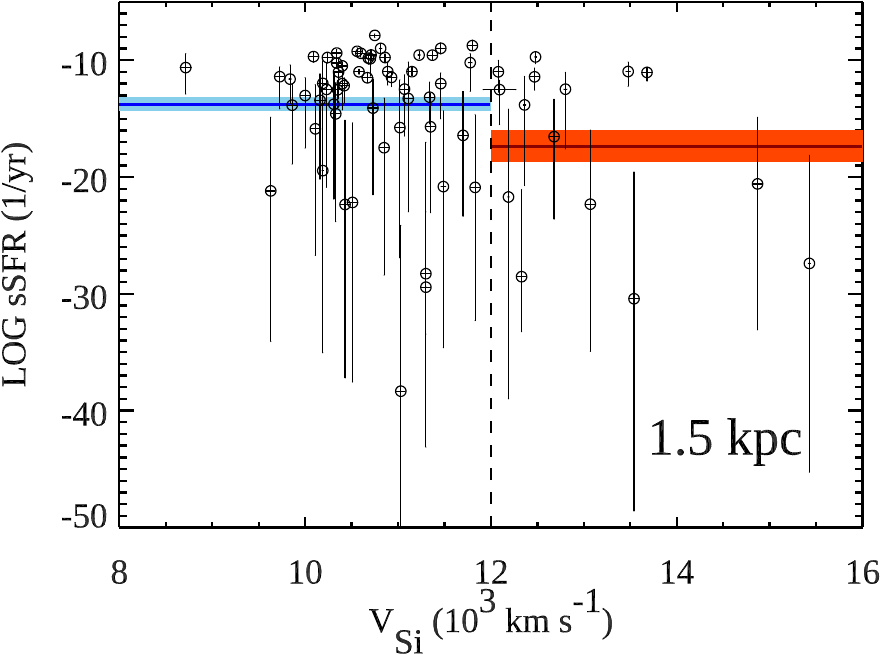}
  \hspace{0.5cm}
  \includegraphics[scale=0.55]{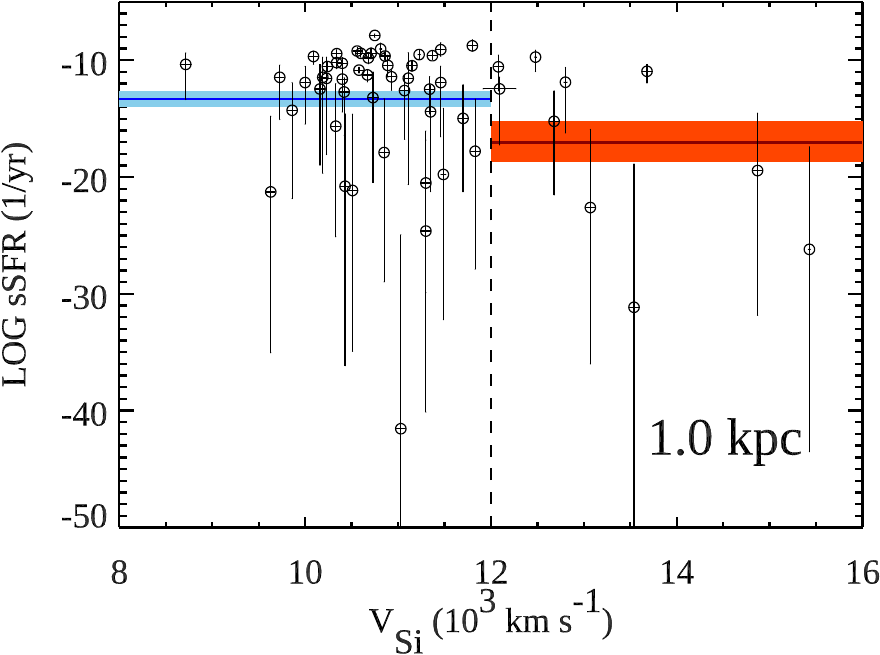}\\
  \vspace{0.6cm}
  \includegraphics[scale=0.55]{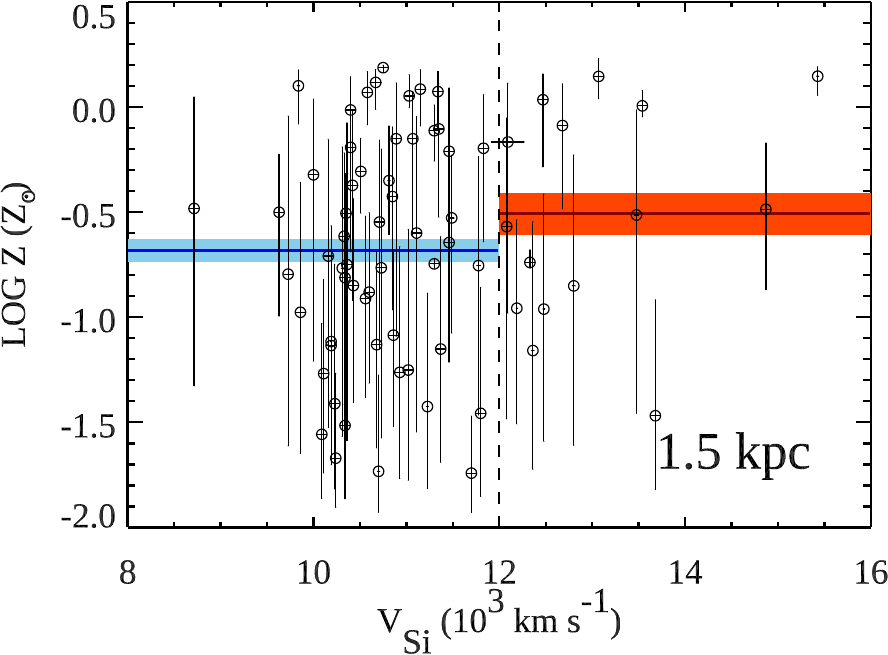}
  \hspace{0.5cm}
  \includegraphics[scale=0.55]{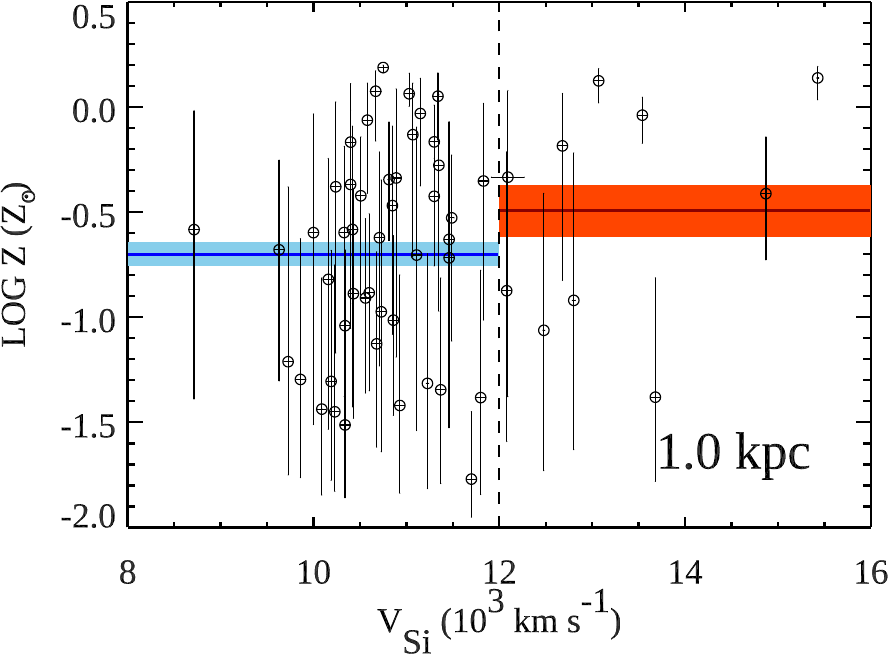}
  \caption{\textit{Left panels}: The host-galaxy mass-weighted stellar age (top), specific star-formation (middle), and mass-weighted stellar metallicity (bottom) measured with an aperture of $r=1.5$\,kpc\ at the SN location, as a function \Siii\ velocity (\vsiii) measured near maximum light. The vertical line represents the criterion used to separate the velocity bins. The solid horizontal line represents the average of each host-galaxy parameter obtained via bootstrap resampling in bins of velocity, and the shaded region shows the $1\sigma$ uncertainty of the mean. \textit{Right panels}: The same as left panels, but with an aperture of $r=1$\,kpc instead.}
  \label{fig:wt34_15kpc}
\end{figure*}

\begin{figure*}
\centering
  \includegraphics[scale=0.38]{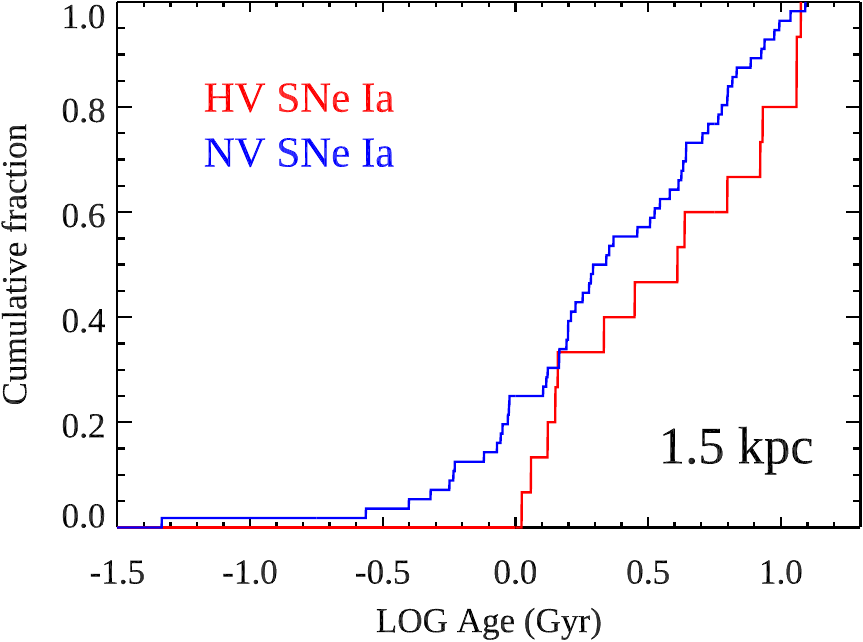}
  \hspace{0.2cm}
  \includegraphics[scale=0.38]{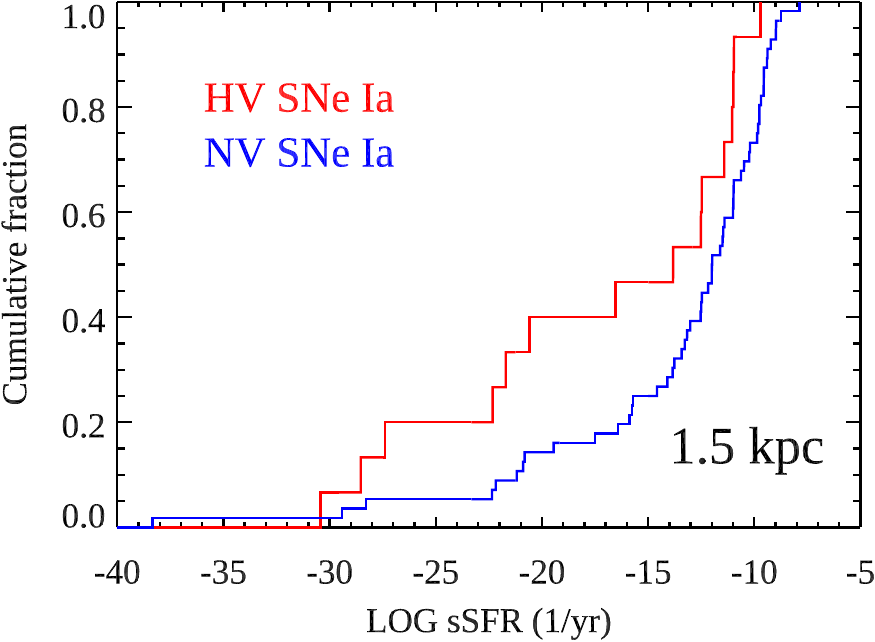}
  \hspace{0.2cm}
  \includegraphics[scale=0.38]{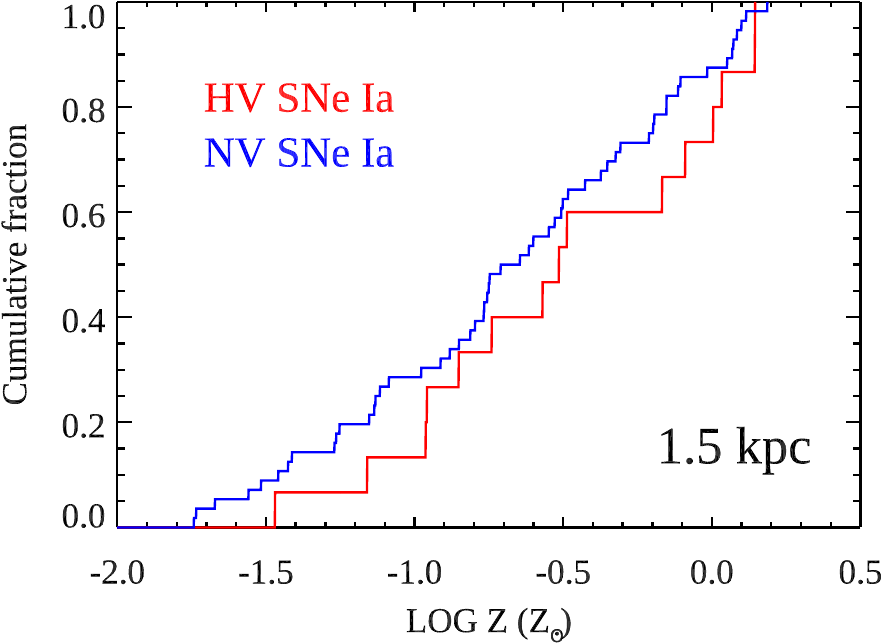}\\
  \vspace{0.2cm}
  \includegraphics[scale=0.38]{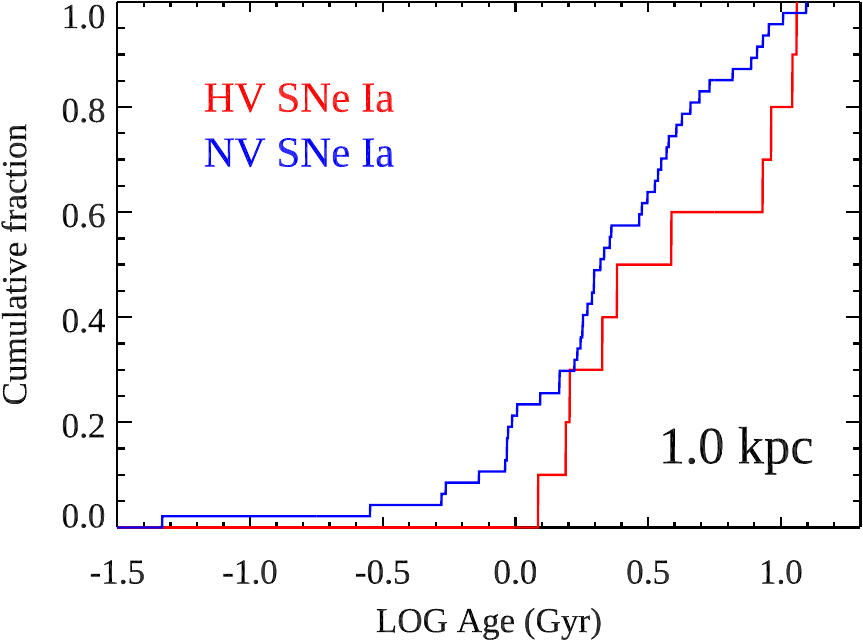}
  \hspace{0.2cm}
  \includegraphics[scale=0.38]{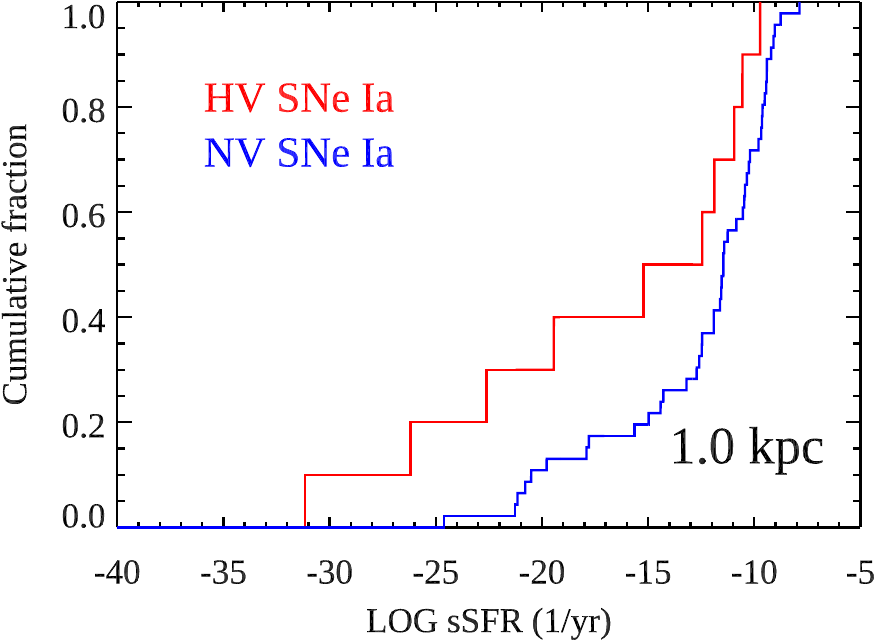}
  \hspace{0.2cm}
  \includegraphics[scale=0.38]{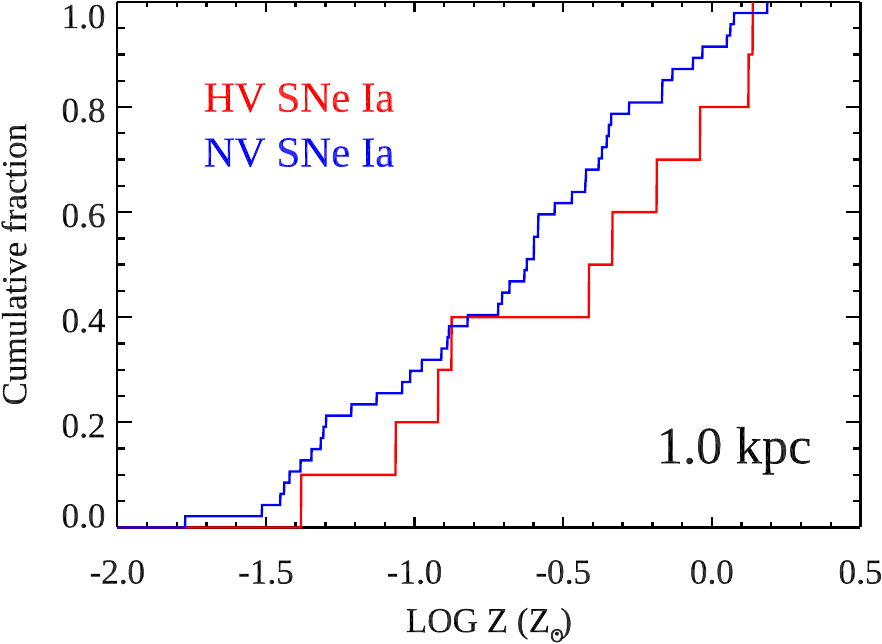}
  \caption{\textit{Upper panels}: The cumulative fractions of host-galaxy mass-weighted stellar age (left), specific star-formation rate (middle), and mass-weighted stellar metallicity (right) measured with an aperture of $r=1.5$\,kpc for HV SNe~Ia (in red) and NV SNe~Ia (in blue), respectively. \textit{Lower panels}: The same as upper panels, but with an aperture of $r=1$\,kpc instead.}
  \label{fig:wt34_15kpc_cdf}
\end{figure*}

\section{Results}
\label{sec:results}
We first compare the properties of the local environment of NV and HV SNe Ia within an aperture radius of 1.5\,kpc. The results are shown in the left panels of Figure~\ref{fig:wt34_15kpc}. Here we calculate a bootstrapping mean instead of a weighted mean as we find the weighted mean is likely biased to some measurements with extremely small uncertainties. A Gini coefficient of 0.9 is derived with the uncertainties of our measurements, indicating that a weighted mean may not be robust enough to reflect the nature of our dataset. 

The difference in mean $\log(\rm age)$ calculated via bootstrap resampling (with a total number of \num{10000} resamplings) between HV and NV subgroups is $0.22\pm0.09$\,Gyr (at $2.4\sigma$ significance). Thus, we only see a potential trend that HV SNe~Ia tend to locate in the environment of older stellar populations than their NV counterparts. 

We also investigate the specific SFR (sSFR) of the local environments for HV and NV SNe~Ia. The sSFR is defined as the SFR per unit \mstellar, which is also a useful indicator for the age of the galaxies, in a sense that higher-sSFR galaxies tend to have younger stellar populations than lower-sSFR galaxies \citep[e.g.,][]{2004MNRAS.351.1151B}. The result from sSFR shows a similar trend, where we determine a difference in mean $\log(\rm sSFR)$ of $3.60\pm1.49$\,yr$^{-1}$ (at $2.4\sigma$ significance), in the sense that the local environments of HV SNe~Ia tend to have lower sSFR than that of NV SNe~Ia. The trend with stellar metallicity is less significant than that of stellar age and sSFR, with HV SNe~Ia having more metal-rich environments than NV counterparts at only $1.5\sigma$ significance. Nevertheless, the stellar metallicity derived in this work (i.e., via broad-band SED fitting) may not be as robust as the stellar age and sSFR \citep[e.g.,][]{2013ARA&A..51..393C}. 

We also compare the results with the local host-galaxy parameters determined with an aperture size of $r=1$\,kpc (see the right panels of Figure~\ref{fig:wt34_15kpc}). Despite a smaller sample size, the trends are generally consistent with those from $r=1.5$\,kpc, with HV SNe~Ia having an older local stellar age and smaller sSFR (at $2.4\sigma$ and $2\sigma$ significance, respectively) than that of NV SNe~Ia. The local environments of HV SNe~Ia are also likely more metal-rich than that of NV counterparts, but at only $1.5\sigma$ significance. We summarize all the statistics in Table~\ref{statistics}.

Figure~\ref{fig:wt34_15kpc_cdf} compares the cumulative fractions of different host-galaxy parameters between NV and HV SNe~Ia. Similar to the results in Figure~\ref{fig:wt34_15kpc}, we find the progenitor systems of HV SNe~Ia could favor older populations compared to those of NV SNe~Ia. Despite our limited sample size, the local environments of all the HV SNe~Ia generally exhibit ages older than $\sim1$\,Gyr, while the distribution of NV SNe~Ia features a tail that extends to $\sim30$\,Myr, irrespective of the aperture radius. However, Kolmogorov-Smirnov (K-S) tests on stellar age, sSFR, and stellar metallicity distributions between HV and NV SNe~Ia all give values of $p>0.2$, indicating that HV and NV SNe~Ia do not exhibit significant differences in term of their host-galaxy distributions. If performing the Anderson–Darling (A-D) test, which is generally more sensitive to the tail of the distribution, we obtain $p$-values of 0.15 and 0.06 (0.14 and 0.09) for stellar age and sSFR determined with an aperture size of $r=1.5$\,kpc ($r=1$\,kpc), respectively. While these numbers are more significant than those from the K-S test, it remains inconclusive to argue that the local host-galaxy environments of HV SNe~Ia are significantly different from those of NV SNe~Ia. 

\begin{table}
\centering
\caption{The bin average of sub-samples shown in Figure~\ref{fig:wt34_15kpc}.}
\begin{tabular}{lccccccc}
\hline\hline
   &  num. of SNe & $\log \rm Age$ & $\log \rm sSFR$ & $\log \rm Z$ \\
   &              &    (Gyr)       & ($\rm yr^{-1}$) & ($\rm Z_{\odot}$) \\
\hline
\multicolumn{5}{c}{$r=1.5$\,kpc}\\
\hline
HV & 15           &$0.56\pm0.08$   & $-17.36\pm1.36$ & $-0.51\pm0.10$ \\
NV & 56           &$0.34\pm0.05$   & $-13.76\pm0.59$ & $-0.68\pm0.06$  \\
\hline
\multicolumn{5}{c}{$r=1.0$\,kpc}\\
\hline
HV & 10           &$0.58\pm0.09$   & $-17.02\pm1.79$ & $-0.49\pm0.12$ \\
NV & 47           &$0.33\pm0.05$   & $-13.31\pm0.69$ & $-0.70\pm0.05$  \\
\hline
\end{tabular}
\label{statistics}
\end{table}

\section{Discussion}
\label{sec:discussion}

\begin{figure*}
	\centering
		\includegraphics*[scale=0.48]{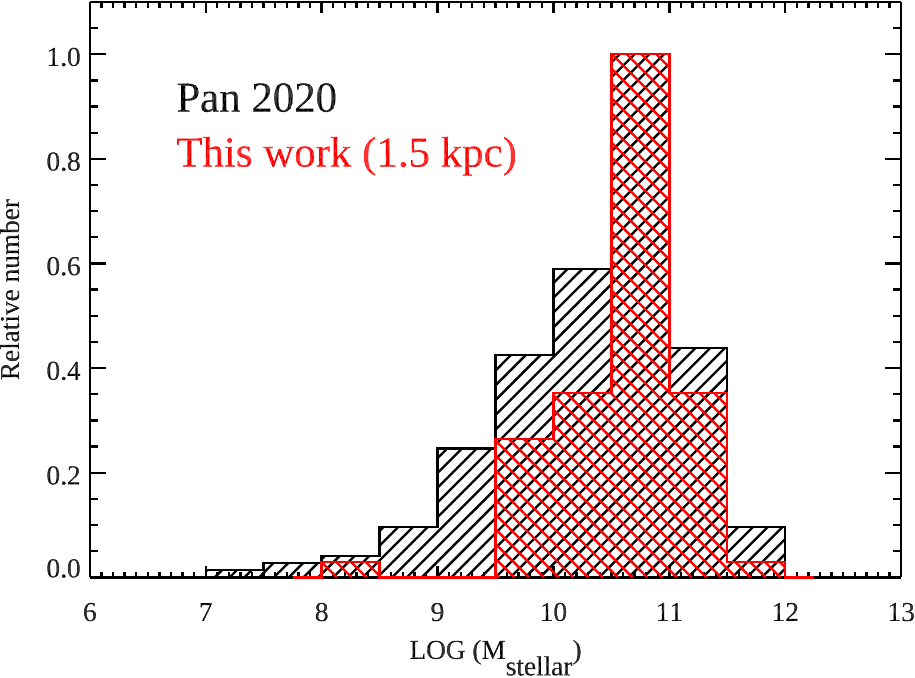}
            \hspace{0.2cm}
  		\includegraphics*[scale=0.48]{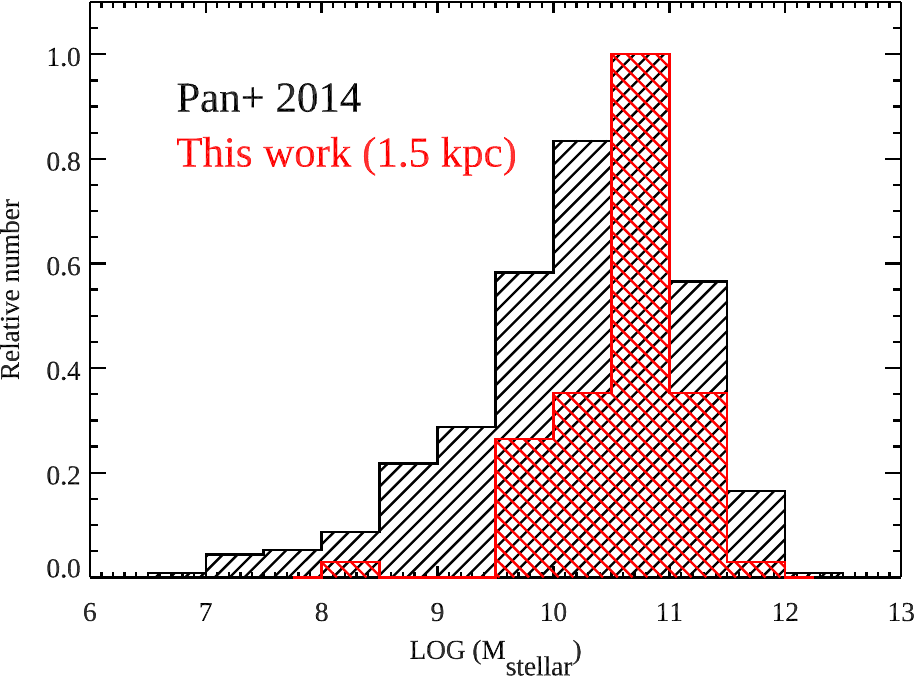}\\
            \vspace{0.2cm}
            \includegraphics*[scale=0.48]{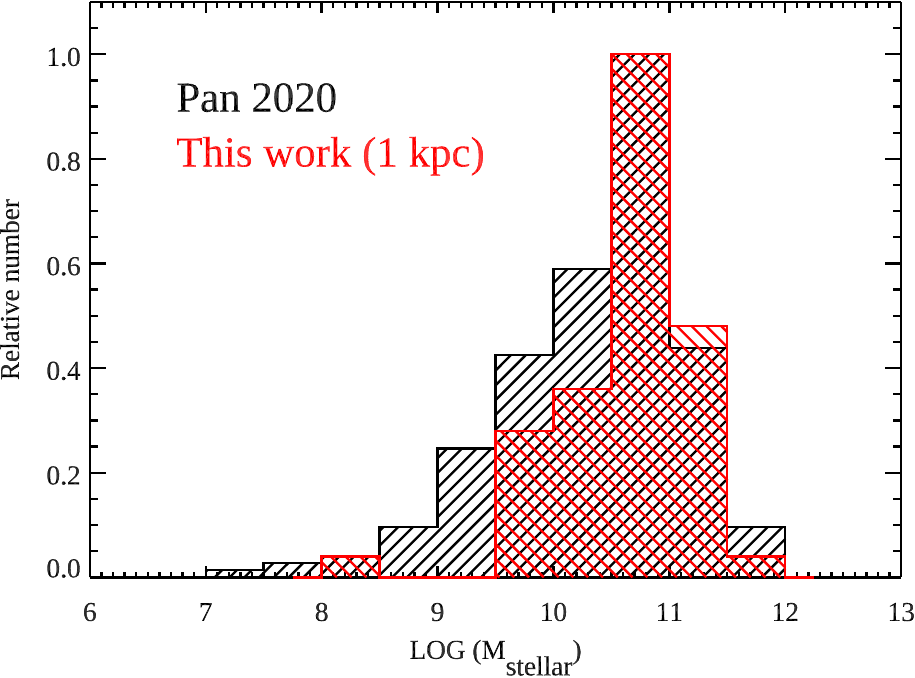}
            \hspace{0.2cm}
  		\includegraphics*[scale=0.48]{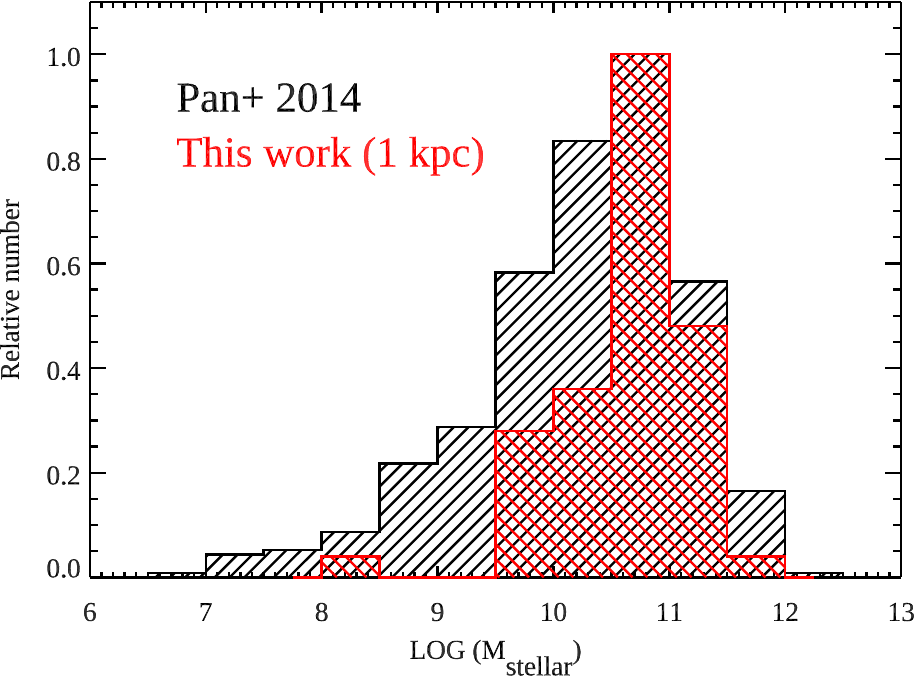}
       \caption{\textit{Left}: The distributions of host-galaxy stellar mass (\mstellar) of this work (red histograms) and \citet[][black histograms]{2020ApJ...895L...5P} for $r=1.5$\,kpc (upper) and 1\,kpc (lower), respectively. \textit{Right}: The same as the left panels, but with a comparison to the pure PTF sample from \citet{2014MNRAS.438.1391P}.
      }
        \label{mass-compare}
\end{figure*}

By studying the global host-galaxy properties of a large low-$z$ sample, \citet{2015MNRAS.446..354P} and \citet{2020ApJ...895L...5P} found that HV SNe~Ia are predominantly discovered in more massive galaxies, while NV SNe~Ia can be found in both low-mass and massive galaxies, suggesting that the progenitor systems of HV SNe~Ia (and possibly some NV SNe~Ia) may originate from relatively old and/or metal-rich populations. This was further supported by the results in \citet{2022arXiv221106895P} that HV SNe~Ia tend to explode at the local Universe compared to their NV counterparts. However, \citet{2020ApJ...895L...5P} did not see a significant difference with global sSFR between HV and NV SNe~Ia in their sample (at only $\sim1\sigma$ significance). These recent studies motivated us to look closely at the local environments of both SN populations.

This work examines the local environments of a sub-sample of nearby NV and HV SNe~Ia studied in \citet{2020ApJ...895L...5P}. Our results reveal only a possible trend (at $\sim2.4\sigma$ significance) between the local environmental properties (such as mass-weighted stellar age and sSFR) of HV and NV SNe~Ia, in the sense that HV SNe~Ia tend to originate from older populations than NV SNe~Ia. Nevertheless, this trend is more significant than that determined by the global sSFR in \citet{2020ApJ...895L...5P}. While our results are inconclusive, they may not support the previous findings that the HV SNe~Ia could be associated with younger stellar populations than their NV counterparts \citep[e.g.,][]{2013Sci...340..170W}. Unfortunately, our measurement of host-galaxy metallicity is less robust than that of age, such that we cannot certainly say whether age or metallicity is a more important factor.

Although it is possible that our HV SN~Ia sample is small, we are aware that the sample studied in this work could exhibit serious bias. The left panels in Figure~\ref{mass-compare} compare the host-galaxy \mstellar\ distributions between our sample and the parent sample studied in \citet{2020ApJ...895L...5P}. We find our study preferentially selects more massive galaxies (i.e., $\log\mstellar\gtrsim10\,M_{\odot}$) from the parent sample. The K-S test gives a $p$-value of 0.06 and 0.02 for $r=1.5$\,kpc and 1\,kpc samples, respectively, that our sample and parent sample are similar populations. This bias becomes more significant if the comparison is made to the pure PTF sample studied in \citet{2014MNRAS.438.1391P}, which can be seen in the right panels of Figure~\ref{mass-compare}. That said, our sample likely has a deficit in low-mass galaxies and does not probe the full parameter space of the parent sample. The low-mass galaxies generally host younger stellar populations \citep[e.g.,][]{2005MNRAS.362...41G} and presumably produce a larger fraction of NV SNe~Ia \citep[e.g.,][]{2020ApJ...895L...5P}. This is probably not surprising given that we have to implement certain redshift cuts (see Secion~\ref{sec:host-data}) on the parent sample to justify the measurement of `local' host-galaxy properties. Thus, we tend to select those more nearby and/or massive galaxies as our targets. The lack of low-mass galaxies in the sample may limit our investigation of the environmental difference between NV and HV SNe~Ia. Higher-resolution images with a broad wavelength coverage will be critical for future SN local environmental studies to probe a more complete parameter space.

It is also evident from recent studies that the velocity alone may not be sufficient for separating different SN~Ia populations \citep{2020MNRAS.499.5325Z,2022arXiv221106895P}. These studies found the \Siii\ velocity shows a significant bimodal Gaussian distribution. Assuming this bimodal distribution results from two physically distinct populations, \citet{2022arXiv221106895P} showed that more than 50\% of SNe~Ia in the `higher mean velocity' population (to differentiate from HV SNe~Ia) of the bimodal distribution are classified as NV SNe~Ia due to the overlap in velocity space, while only $\sim3$\% of SNe~Ia in the `lower mean velocity' population (to differentiate from NV SNe~Ia) are classified as HV SNe~Ia. The overlap between two underlying populations in velocity space may reduce the significance of differences (if any) between HV and NV SN~Ia host-galaxy environments. For example, some NV SNe~Ia may be produced in environments similar to those HV SNe~Ia (and vice versa). 

Previous studies proposed that the double detonation from sub-Chandrasekhar-mass WD could be a promising mechanism to explain the observational characteristics of HV SNe~Ia (and potentially part of the NV SNe~Ia), including their high \vsiii, unique color, and asymmetric explosions \citep[e.g.,][]{2019ApJ...873...84P,2021ApJ...922...68S,2021ApJ...906...99L}. While our results are not conclusive, the preference of old stellar populations might imply that the progenitors of HV SNe~Ia could be associated with old stellar populations and favor scenarios of long delay time (i.e., the time between the progenitor star formation and the subsequent SN~Ia explosion), such as some double-degenerate scenarios. This is also supported by recent theoretical studies indicating that the double detonation of the primary WD could be ignited by accreting a thin helium shell from a helium WD donor during the dynamical phase of the merger \citep[e.g.,][]{2018ApJ...854...52S,2018ApJ...865...15S,2019ApJ...878L..38T,2022MNRAS.517.5260P}. However, it is worth noting that this process could also be triggered in single-degenerate scenarios \citep[e.g., a WD plus a helium star;][]{2021ApJ...906...99L}. Thus, the age of the progenitor could put strong constraints on the SN delay time and further distinguish between different channels.

Nevertheless, we do not rule out the possibility that age may not be the major (or the only) factor driving the difference between HV and NV SN~Ia populations. As mentioned above, the host-galaxy metallicity measured in this work is not as robust as age and sSFR. Studies with local host-galaxy spectroscopic analysis may provide us with more information in disentangling the differences between SN~Ia populations (Jim\'eniz~et~al., in preparation). 

We also note the existing degeneracy between age and metallicity in SED fitting that could complicate the inference of these parameters in our analysis. However, the imaging data used in this work covers a wide range of wavelengths (from rest-frame UV to NIR), which can help break such degeneracy. For instance, previous studies have shown that SED fitting with photometric data that covers the rest-frame UV-to-NIR can break the degeneracy between age and metallicity (possibly even dust) because the reddening effects by these parameters are distinguishable by the color sets of the data \citep[e.g.,][]{2019ApJ...877..141M,2023ApJ...945..117A}. Moreover, multiple tests have been performed on \textsc{piXedfit} by \citet{2021ApJS..254...15A,2022ApJ...926...81A,2023ApJ...945..117A}, showing its capabilities in robustly inferring stellar population properties, including stellar mass, SFR, age, metallicity, and dust attenuation level. One such test using mock photometric data generated from TNG simulation shows that \textsc{piXedfit} can recover the stellar population properties reasonably well when the data cover at least the rest-frame UV to NIR. The mock photometry used in that test are also from the same set of filters as those in our analysis.

\section{Conclusions}
\label{sec:conclusions}

This study examines the local host-galaxy properties at the SN position and explores the potential environmental differences between HV and NV SNe~Ia. We only see a potential trend (at $2.4\sigma$ significance) that HV SNe~Ia may be associated with older stellar populations than their NV counterparts. No significant trend is identified with stellar metallicity, although our broad-band photometric data hinder a robust metallicity measurement. While it is possible that the host-galaxy age alone might not be the underlying factor driving the differences between NV and HV SNe~Ia, we note that our sample is likely biased and unable to probe the diversities of SN~Ia host-galaxy environments. There is also likely a degeneracy between the underlying populations of HV and NV SNe~Ia. Future studies with less biased samples and investigation into more properties will be crucial to unveiling the nature of HV SNe~Ia.

\section*{Acknowledgements}
Y.-C.P. is supported by the National Science and Technology Council (NSTC grant 112-2112-M-008-026-MY3).
%maybe add acknowledgement for the catalogs

%%%%%%%%%%%%%%%%%%%%%%%%%%%%%%%%%%%%%%%%%%%%%%%%%%
\section*{Data Availability}
The data underlying this article will be shared on reasonable request to the corresponding author.
% The inclusion of a Data Availability Statement is a requirement for articles published in MNRAS. Data Availability Statements provide a standardised format for readers to understand the availability of data underlying the research results described in the article. The statement may refer to original data generated in the course of the study or to third-party data analysed in the article. The statement should describe and provide means of access, where possible, by linking to the data or providing the required accession numbers for the relevant databases or DOIs.
%%%%%%%%%%%%%%%%%%%% REFERENCES %%%%%%%%%%%%%%%%%%

% The best way to enter references is to use BibTeX:

\bibliographystyle{mnras}
\bibliography{snia_local} % if your bibtex file is called example.bib

% Alternatively you could enter them by hand, like this:
% This method is tedious and prone to error if you have lots of references
%\begin{thebibliography}{99}
%\bibitem[\protect\citeauthoryear{Author}{2012}]{Author2012}
%Author A.~N., 2013, Journal of Improbable Astronomy, 1, 1
%\bibitem[\protect\citeauthoryear{Others}{2013}]{Others2013}
%Others S., 2012, Journal of Interesting Stuff, 17, 198
%\end{thebibliography}

%%%%%%%%%%%%%%%%%%%%%%%%%%%%%%%%%%%%%%%%%%%%%%%%%%

%%%%%%%%%%%%%%%%% APPENDICES %%%%%%%%%%%%%%%%%%%%%

\appendix
\section{The SN~Ia host-galaxy properties measured in this work}

\begin{table*}
\centering
\caption{Host-galaxy properties of our $r=1.5$\,kpc SN sample measured in this work.}
\renewcommand{\arraystretch}{1.5}
%\scriptsize
\begin{tabular}{lrrr}
\hline\hline
SN Name         & $\log \rm Age$ & $\log \rm sSFR$ & $\log \rm Z$ \\
                &    (Gyr)       & ($\rm yr^{-1}$) & ($\rm Z_{\odot}$) \\
\hline
PTF09dnp & $ 0.93 ^{+ 0.12 }_{- 0.18 }$ & $ -21.71 ^{+ 7.52 }_{- 17.28 }$ & $ -0.96 ^{+ 0.42 }_{- 0.55 }$ \\
PTF09foz & $ 0.89 ^{+ 0.14 }_{- 0.22 }$ & $ -19.45 ^{+ 5.89 }_{- 15.59 }$ & $ -1.12 ^{+ 0.39 }_{- 0.54 }$ \\
PTF10fxl & $ 0.28 ^{+ 0.27 }_{- 0.32 }$ & $ -9.55 ^{+ 0.36 }_{- 0.30 }$ & $ -1.42 ^{+ 0.54 }_{- 0.39 }$ \\
PTF10hld & $ 0.02 ^{+ 0.34 }_{- 0.15 }$ & $ -10.96 ^{+ 0.81 }_{- 1.28 }$ & $ -0.51 ^{+ 0.5 }_{- 0.94 }$ \\
PTF10lot & $ 0.33 ^{+ 0.42 }_{- 0.29 }$ & $ -12.47 ^{+ 1.44 }_{- 5.11 }$ & $ -0.85 ^{+ 0.62 }_{- 0.76 }$ \\
PTF10qhp & $ 0.94 ^{+ 0.11 }_{- 0.18 }$ & $ -20.82 ^{+ 6.49 }_{- 13.79 }$ & $ -0.53 ^{+ 0.3 }_{- 0.55 }$ \\
PTF10qjq & $ -0.40 ^{+ 0.58 }_{- 0.39 }$ & $ -8.98 ^{+ 0.44 }_{- 0.52 }$ & $ -0.35 ^{+ 0.26 }_{- 0.26 }$ \\
PTF10tce & $ 0.64 ^{+ 0.26 }_{- 0.31 }$ & $ -13.82 ^{+ 2.44 }_{- 6.90 }$ & $ -1.16 ^{+ 0.61 }_{- 0.57 }$ \\
PTF10viq & $ 0.15 ^{+ 0.38 }_{- 0.37 }$ & $ -9.71 ^{+ 0.46 }_{- 0.58 }$ & $ -0.96 ^{+ 0.55 }_{- 0.63 }$ \\
PTF10ygu & $ 1.06 ^{+ 0.04 }_{- 0.05 }$ & $ -27.40 ^{+ 9.26 }_{- 17.87 }$ & $ 0.15 ^{+ 0.04 }_{- 0.09 }$ \\
PTF10yux & $ 0.37 ^{+ 0.31 }_{- 0.43 }$ & $ -10.20 ^{+ 0.78 }_{- 2.44 }$ & $ -0.75 ^{+ 0.52 }_{- 0.71 }$ \\
PTF10zai & $ 0.19 ^{+ 0.30 }_{- 0.27 }$ & $ -10.62 ^{+ 1.18 }_{- 2.27 }$ & $ -0.48 ^{+ 0.53 }_{- 0.84 }$ \\
PTF11apk & $ 0.64 ^{+ 0.28 }_{- 0.33 }$ & $ -13.76 ^{+ 2.67 }_{- 8.13 }$ & $ -0.77 ^{+ 0.58 }_{- 0.8 }$ \\
PTF11gdh & $ 0.17 ^{+ 0.14 }_{- 0.26 }$ & $ -11.61 ^{+ 1.22 }_{- 2.10 }$ & $ 0.1 ^{+ 0.08 }_{- 0.18 }$ \\
PTF12awi & $ 0.12 ^{+ 0.48 }_{- 0.18 }$ & $ -11.40 ^{+ 0.84 }_{- 2.74 }$ & $ -0.8 ^{+ 0.75 }_{- 0.82 }$ \\
PTF12dhb & $ -0.23 ^{+ 0.32 }_{- 0.16 }$ & $ -9.87 ^{+ 0.51 }_{- 1.48 }$ & $ -1.73 ^{+ 0.46 }_{- 0.19 }$ \\
PTF12giy & $ 0.16 ^{+ 0.40 }_{- 0.13 }$ & $ -12.51 ^{+ 0.80 }_{- 2.99 }$ & $ -0.17 ^{+ 0.28 }_{- 0.81 }$ \\
SN1989M & $ 0.80 ^{+ 0.01 }_{- 0.01 }$ & $ -28.53 ^{+ 7.46 }_{- 4.71 }$ & $ -0.74 ^{+ 0.06 }_{- 0.03 }$ \\
SN1994S & $ 0.21 ^{+ 0.21 }_{- 0.19 }$ & $ -12.01 ^{+ 1.04 }_{- 2.24 }$ & $ -0.01 ^{+ 0.16 }_{- 0.53 }$ \\
SN1998es & $ -0.12 ^{+ 0.35 }_{- 0.19 }$ & $ -9.76 ^{+ 0.34 }_{- 0.47 }$ & $ -1.67 ^{+ 0.4 }_{- 0.23 }$ \\
SN1999aa & $ 0.12 ^{+ 0.19 }_{- 0.10 }$ & $ -12.53 ^{+ 0.81 }_{- 1.58 }$ & $ -0.51 ^{+ 0.19 }_{- 0.44 }$ \\
SN1999ac & $ 0.25 ^{+ 0.38 }_{- 0.21 }$ & $ -10.48 ^{+ 0.32 }_{- 0.33 }$ & $ -0.19 ^{+ 0.26 }_{- 0.5 }$ \\
SN1999dq & $ 0.29 ^{+ 0.31 }_{- 0.31 }$ & $ -9.76 ^{+ 0.33 }_{- 0.39 }$ & $ -1.09 ^{+ 0.4 }_{- 0.43 }$ \\
SN1999gd & $ 0.11 ^{+ 0.25 }_{- 0.12 }$ & $ -12.17 ^{+ 0.68 }_{- 1.50 }$ & $ -0.37 ^{+ 0.36 }_{- 0.55 }$ \\
SN2000dn & $ 0.55 ^{+ 0.31 }_{- 0.39 }$ & $ -11.98 ^{+ 1.95 }_{- 7.24 }$ & $ -1.13 ^{+ 0.57 }_{- 0.57 }$ \\
SN2001da & $ 0.63 ^{+ 0.25 }_{- 0.22 }$ & $ -15.70 ^{+ 2.85 }_{- 7.35 }$ & $ -0.1 ^{+ 0.21 }_{- 0.42 }$ \\
SN2001fe & $ 0.34 ^{+ 0.36 }_{- 0.24 }$ & $ -12.47 ^{+ 1.23 }_{- 4.02 }$ & $ -0.15 ^{+ 0.23 }_{- 0.46 }$ \\
SN2002bf & $ 0.45 ^{+ 0.33 }_{- 0.31 }$ & $ -11.04 ^{+ 0.46 }_{- 0.75 }$ & $ -1.47 ^{+ 0.55 }_{- 0.35 }$ \\
SN2002bo & $ 1.08 ^{+ 0.18 }_{- 0.06 }$ & $ -22.32 ^{+ 6.34 }_{- 12.61 }$ & $ 0.15 ^{+ 0.09 }_{- 0.11 }$ \\
SN2002eb & $ 0.63 ^{+ 0.29 }_{- 0.39 }$ & $ -12.49 ^{+ 2.24 }_{- 8.38 }$ & $ -1.41 ^{+ 0.66 }_{- 0.41 }$ \\
SN2002eu & $ 0.78 ^{+ 0.21 }_{- 0.33 }$ & $ -15.76 ^{+ 4.08 }_{- 11.16 }$ & $ -1.25 ^{+ 0.67 }_{- 0.53 }$ \\
SN2002ha & $ 0.46 ^{+ 0.36 }_{- 0.30 }$ & $ -11.46 ^{+ 0.52 }_{- 0.79 }$ & $ -1.26 ^{+ 0.6 }_{- 0.51 }$ \\
SN2003U & $ 0.80 ^{+ 0.01 }_{- 0.01 }$ & $ -29.43 ^{+ 6.45 }_{- 4.02 }$ & $ -0.75 ^{+ 0.03 }_{- 0.02 }$ \\
SN2003Y & $ 0.51 ^{+ 0.30 }_{- 0.23 }$ & $ -13.85 ^{+ 1.84 }_{- 5.04 }$ & $ -0.98 ^{+ 0.62 }_{- 0.67 }$ \\
SN2003cq & $ 0.12 ^{+ 0.38 }_{- 0.22 }$ & $ -10.98 ^{+ 0.97 }_{- 1.75 }$ & $ -0.57 ^{+ 0.52 }_{- 0.91 }$ \\
SN2004dt & $ 1.06 ^{+ 0.04 }_{- 0.07 }$ & $ -30.42 ^{+ 10.86 }_{- 18.18 }$ & $ 0.01 ^{+ 0.07 }_{- 0.05 }$ \\
SN2004gs & $ 0.93 ^{+ 0.12 }_{- 0.16 }$ & $ -22.35 ^{+ 7.22 }_{- 14.85 }$ & $ -0.85 ^{+ 0.41 }_{- 0.56 }$ \\
SN2005M & $ -0.02 ^{+ 0.05 }_{- 0.05 }$ & $ -11.49 ^{+ 0.47 }_{- 0.41 }$ & $ 0.12 ^{+ 0.06 }_{- 0.13 }$ \\
SN2005ao & $ -0.23 ^{+ 0.43 }_{- 0.45 }$ & $ -8.97 ^{+ 0.48 }_{- 0.49 }$ & $ -0.64 ^{+ 0.4 }_{- 0.57 }$ \\
SN2005eq & $ -0.03 ^{+ 0.44 }_{- 0.31 }$ & $ -9.68 ^{+ 0.38 }_{- 0.49 }$ & $ -1.56 ^{+ 0.53 }_{- 0.31 }$ \\
SN2005ki & $ 1.09 ^{+ 0.01 }_{- 0.02 }$ & $ -38.34 ^{+ 14.19 }_{- 16.86 }$ & $ 0.05 ^{+ 0.1 }_{- 0.06 }$ \\
\hline
\label{sample_1.5_1}
\end{tabular}
\end{table*}

\begin{table*}
\centering
\caption{Host-galaxy properties of our $r=1.5$\,kpc SN sample measured in this work (continued).}
\renewcommand{\arraystretch}{1.5}
%\scriptsize
\begin{tabular}{lrrr}
\hline\hline
SN Name         & $\log \rm Age$ & $\log \rm sSFR$ & $\log \rm Z$ \\
                &    (Gyr)       & ($\rm yr^{-1}$) & ($\rm Z_{\odot}$) \\
\hline
SN2006N & $ 1.04 ^{+ 0.06 }_{- 0.11 }$ & $ -28.28 ^{+ 11.25 }_{- 14.84 }$ & $ -0.11 ^{+ 0.12 }_{- 0.14 }$ \\
SN2006S & $ 0.23 ^{+ 0.34 }_{- 0.47 }$ & $ -9.55 ^{+ 0.47 }_{- 0.40 }$ & $ -0.55 ^{+ 0.39 }_{- 0.61 }$ \\
SN2006bt & $ 1.00 ^{+ 0.07 }_{- 0.11 }$ & $ -22.17 ^{+ 6.82 }_{- 15.38 }$ & $ -0.31 ^{+ 0.16 }_{- 0.2 }$ \\
SN2006bz & $ 0.83 ^{+ 0.17 }_{- 0.24 }$ & $ -17.49 ^{+ 4.26 }_{- 10.91 }$ & $ -0.43 ^{+ 0.33 }_{- 0.54 }$ \\
SN2006cm & $ 0.71 ^{+ 0.17 }_{- 0.24 }$ & $ -10.97 ^{+ 0.44 }_{- 0.35 }$ & $ 0.09 ^{+ 0.09 }_{- 0.18 }$ \\
SN2006cq & $ 0.53 ^{+ 0.33 }_{- 0.32 }$ & $ -13.42 ^{+ 2.27 }_{- 6.77 }$ & $ -0.71 ^{+ 0.55 }_{- 0.82 }$ \\
SN2006cs & $ 0.64 ^{+ 0.27 }_{- 0.30 }$ & $ -14.09 ^{+ 2.53 }_{- 7.43 }$ & $ -0.76 ^{+ 0.56 }_{- 0.81 }$ \\
SN2006or & $ 0.29 ^{+ 0.16 }_{- 0.12 }$ & $ -13.16 ^{+ 1.29 }_{- 2.28 }$ & $ 0.07 ^{+ 0.1 }_{- 0.18 }$ \\
SN2006sr & $ 0.06 ^{+ 0.15 }_{- 0.12 }$ & $ -11.41 ^{+ 0.75 }_{- 1.13 }$ & $ 0.03 ^{+ 0.12 }_{- 0.32 }$ \\
SN2007A & $ 0.17 ^{+ 0.30 }_{- 0.33 }$ & $ -9.40 ^{+ 0.37 }_{- 0.34 }$ & $ -0.88 ^{+ 0.38 }_{- 0.43 }$ \\
SN2007N & $ 0.77 ^{+ 0.20 }_{- 0.27 }$ & $ -14.58 ^{+ 2.66 }_{- 9.24 }$ & $ -0.61 ^{+ 0.4 }_{- 0.8 }$ \\
SN2007O & $ 0.35 ^{+ 0.42 }_{- 0.21 }$ & $ -13.02 ^{+ 1.53 }_{- 4.50 }$ & $ -0.32 ^{+ 0.36 }_{- 0.89 }$ \\
SN2007af & $ -0.32 ^{+ 0.38 }_{- 0.24 }$ & $ -9.23 ^{+ 0.31 }_{- 0.39 }$ & $ -0.91 ^{+ 0.39 }_{- 0.47 }$ \\
SN2007ba & $ 0.98 ^{+ 0.08 }_{- 0.15 }$ & $ -21.18 ^{+ 6.32 }_{- 12.89 }$ & $ -0.5 ^{+ 0.27 }_{- 0.49 }$ \\
SN2007bz & $ 0.73 ^{+ 0.19 }_{- 0.16 }$ & $ -16.43 ^{+ 3.76 }_{- 6.93 }$ & $ -1.74 ^{+ 0.27 }_{- 0.19 }$ \\
SN2007ci & $ 0.82 ^{+ 0.17 }_{- 0.21 }$ & $ -20.88 ^{+ 6.24 }_{- 11.41 }$ & $ -0.2 ^{+ 0.25 }_{- 0.44 }$ \\
SN2007gi & $ 0.92 ^{+ 0.12 }_{- 0.18 }$ & $ -20.58 ^{+ 5.70 }_{- 12.49 }$ & $ -0.49 ^{+ 0.32 }_{- 0.38 }$ \\
SN2007s & $ -0.05 ^{+ 0.49 }_{- 0.33 }$ & $ -9.56 ^{+ 0.36 }_{- 0.41 }$ & $ -1.15 ^{+ 0.53 }_{- 0.54 }$ \\
SN2008ar & $ 0.58 ^{+ 0.19 }_{- 0.40 }$ & $ -10.24 ^{+ 0.46 }_{- 0.46 }$ & $ -1.52 ^{+ 0.53 }_{- 0.35 }$ \\
SN2008ec & $ -1.33 ^{+ 0.15 }_{- 0.10 }$ & $ -7.87 ^{+ 0.13 }_{- 0.18 }$ & $ 0.19 ^{+ 0.01 }_{- 0.02 }$ \\
SN2009an & $ 0.61 ^{+ 0.25 }_{- 0.20 }$ & $ -16.54 ^{+ 3.20 }_{- 7.07 }$ & $ -0.09 ^{+ 0.2 }_{- 0.4 }$ \\
SN2009fx & $ 0.80 ^{+ 0.19 }_{- 0.26 }$ & $ -15.87 ^{+ 3.77 }_{- 10.84 }$ & $ -1.27 ^{+ 0.45 }_{- 0.47 }$ \\
SN2010ex & $ -0.02 ^{+ 0.20 }_{- 0.12 }$ & $ -10.97 ^{+ 0.78 }_{- 1.15 }$ & $ -0.15 ^{+ 0.27 }_{- 0.94 }$ \\
SN2010iw & $ 0.20 ^{+ 0.41 }_{- 0.25 }$ & $ -11.00 ^{+ 1.16 }_{- 2.30 }$ & $ -0.75 ^{+ 0.68 }_{- 0.84 }$ \\
SN2011ao & $ -0.25 ^{+ 0.43 }_{- 0.26 }$ & $ -9.38 ^{+ 0.32 }_{- 0.41 }$ & $ -0.81 ^{+ 0.35 }_{- 0.48 }$ \\
SN2011hb & $ -0.56 ^{+ 0.28 }_{- 0.37 }$ & $ -8.75 ^{+ 0.48 }_{- 0.34 }$ & $ -1.46 ^{+ 0.6 }_{- 0.4 }$ \\
SN2011ia & $ -0.07 ^{+ 0.48 }_{- 0.26 }$ & $ -9.82 ^{+ 0.35 }_{- 0.52 }$ & $ -1.13 ^{+ 0.44 }_{- 0.49 }$ \\
SN2012cg & $ -0.05 ^{+ 0.08 }_{- 0.07 }$ & $ -10.99 ^{+ 0.22 }_{- 0.23 }$ & $ 0.07 ^{+ 0.1 }_{- 0.15 }$ \\
SN2012da & $ 0.62 ^{+ 0.31 }_{- 0.39 }$ & $ -13.27 ^{+ 3.12 }_{- 9.70 }$ & $ -0.6 ^{+ 0.55 }_{- 0.95 }$ \\
SN2013di & $ 0.20 ^{+ 0.41 }_{- 0.18 }$ & $ -12.00 ^{+ 0.91 }_{- 3.02 }$ & $ -0.21 ^{+ 0.3 }_{- 0.75 }$ \\
\hline
\label{sample_1.5_2}
\end{tabular}
\end{table*}

\begin{table*}
\centering
\caption{Host-galaxy properties of our $r=1.0$\,kpc SN sample measured in this work.}
\renewcommand{\arraystretch}{1.5}
%\scriptsize
\begin{tabular}{lrrr}
\hline\hline
SN Name         & $\log \rm Age$ & $\log \rm sSFR$ & $\log \rm Z$ \\
                &    (Gyr)       & ($\rm yr^{-1}$) & ($\rm Z_{\odot}$) \\
\hline
PTF10fxl & $ 0.25 ^{+ 0.30 }_{- 0.39 }$ & $ -9.51 ^{+ 0.45 }_{- 0.36 }$ & $ -1.31 ^{+ 0.62 }_{- 0.50 }$ \\
PTF10lot & $ 0.33 ^{+ 0.39 }_{- 0.30 }$ & $ -11.88 ^{+ 1.27 }_{- 4.34 }$ & $ -0.92 ^{+ 0.70 }_{- 0.71 }$ \\
PTF10qhp & $ 0.91 ^{+ 0.12 }_{- 0.18 }$ & $ -19.78 ^{+ 5.65 }_{- 12.44 }$ & $ -0.53 ^{+ 0.30 }_{- 0.59 }$ \\
PTF10qjq & $ -0.28 ^{+ 0.52 }_{- 0.49 }$ & $ -9.03 ^{+ 0.51 }_{- 0.53 }$ & $ -0.35 ^{+ 0.27 }_{- 0.29 }$ \\
PTF10viq & $ 0.09 ^{+ 0.44 }_{- 0.40 }$ & $ -9.73 ^{+ 0.58 }_{- 1.23 }$ & $ -1.06 ^{+ 0.65 }_{- 0.67 }$ \\
PTF10ygu & $ 1.04 ^{+ 0.06 }_{- 0.07 }$ & $ -26.19 ^{+ 8.79 }_{- 17.33 }$ & $ 0.14 ^{+ 0.05 }_{- 0.10 }$ \\
PTF10zai & $ 0.26 ^{+ 0.38 }_{- 0.36 }$ & $ -10.35 ^{+ 0.97 }_{- 3.01 }$ & $ -0.58 ^{+ 0.57 }_{- 0.81 }$ \\
PTF12awi & $ 0.32 ^{+ 0.36 }_{- 0.32 }$ & $ -11.46 ^{+ 1.04 }_{- 3.61 }$ & $ -1.21 ^{+ 0.83 }_{- 0.54 }$ \\
PTF12giy & $ 0.21 ^{+ 0.55 }_{- 0.20 }$ & $ -12.45 ^{+ 1.01 }_{- 4.79 }$ & $ -0.33 ^{+ 0.41 }_{- 1.04 }$ \\
SN1994S & $ 0.23 ^{+ 0.33 }_{- 0.23 }$ & $ -11.61 ^{+ 1.28 }_{- 2.81 }$ & $ -0.17 ^{+ 0.28 }_{- 0.89 }$ \\
SN1998es & $ -0.01 ^{+ 0.22 }_{- 0.16 }$ & $ -10.53 ^{+ 0.45 }_{- 0.44 }$ & $ -0.38 ^{+ 0.40 }_{- 0.79 }$ \\
SN1999ac & $ 0.27 ^{+ 0.35 }_{- 0.27 }$ & $ -10.25 ^{+ 0.49 }_{- 0.41 }$ & $ -0.37 ^{+ 0.35 }_{- 0.57 }$ \\
SN1999dq & $ 0.30 ^{+ 0.29 }_{- 0.36 }$ & $ -9.63 ^{+ 0.36 }_{- 0.36 }$ & $ -1.01 ^{+ 0.40 }_{- 0.45 }$ \\
SN1999gd & $ 0.22 ^{+ 0.36 }_{- 0.20 }$ & $ -12.71 ^{+ 1.22 }_{- 3.71 }$ & $ -0.58 ^{+ 0.49 }_{- 0.84 }$ \\
SN2000dn & $ 0.47 ^{+ 0.37 }_{- 0.43 }$ & $ -11.46 ^{+ 1.73 }_{- 8.23 }$ & $ -1.31 ^{+ 0.62 }_{- 0.47 }$ \\
SN2001da & $ 0.61 ^{+ 0.27 }_{- 0.28 }$ & $ -14.41 ^{+ 2.29 }_{- 6.84 }$ & $ -0.28 ^{+ 0.34 }_{- 0.69 }$ \\
SN2001fe & $ 0.30 ^{+ 0.42 }_{- 0.19 }$ & $ -12.59 ^{+ 1.32 }_{- 4.18 }$ & $ -0.13 ^{+ 0.25 }_{- 0.58 }$ \\
SN2002bf & $ 0.38 ^{+ 0.35 }_{- 0.29 }$ & $ -10.94 ^{+ 0.59 }_{- 1.01 }$ & $ -1.38 ^{+ 0.57 }_{- 0.40 }$ \\
SN2002bo & $ 0.96 ^{+ 0.10 }_{- 0.14 }$ & $ -22.61 ^{+ 6.69 }_{- 13.39 }$ & $ 0.12 ^{+ 0.06 }_{- 0.10 }$ \\
SN2002eb & $ 0.53 ^{+ 0.35 }_{- 0.43 }$ & $ -11.55 ^{+ 1.82 }_{- 6.51 }$ & $ -1.45 ^{+ 0.70 }_{- 0.38 }$ \\
SN2002ha & $ 0.36 ^{+ 0.40 }_{- 0.24 }$ & $ -11.42 ^{+ 0.58 }_{- 1.15 }$ & $ -1.42 ^{+ 0.62 }_{- 0.42 }$ \\
SN2003U & $ 0.58 ^{+ 0.23 }_{- 0.05 }$ & $ -20.51 ^{+ 3.67 }_{- 9.39 }$ & $ -0.42 ^{+ 0.18 }_{- 0.33 }$ \\
SN2003Y & $ 0.57 ^{+ 0.31 }_{- 0.28 }$ & $ -14.28 ^{+ 2.37 }_{- 7.57 }$ & $ -1.30 ^{+ 0.67 }_{- 0.47 }$ \\
SN2003cq & $ 0.19 ^{+ 0.43 }_{- 0.28 }$ & $ -10.56 ^{+ 0.99 }_{- 2.06 }$ & $ -0.87 ^{+ 0.66 }_{- 0.72 }$ \\
SN2004dt & $ 1.06 ^{+ 0.24 }_{- 0.09 }$ & $ -31.15 ^{+ 12.26 }_{- 20.65 }$ & $ -0.04 ^{+ 0.09 }_{- 0.13 }$ \\
SN2004gs & $ 0.89 ^{+ 0.15 }_{- 0.19 }$ & $ -20.78 ^{+ 6.21 }_{- 15.37 }$ & $ -0.89 ^{+ 0.50 }_{- 0.59 }$ \\
SN2005M & $ -0.04 ^{+ 0.06 }_{- 0.07 }$ & $ -11.25 ^{+ 0.47 }_{- 0.60 }$ & $ 0.08 ^{+ 0.10 }_{- 0.24 }$ \\
SN2005ao & $ -0.14 ^{+ 0.48 }_{- 0.56 }$ & $ -9.09 ^{+ 0.60 }_{- 0.55 }$ & $ -0.63 ^{+ 0.45 }_{- 0.62 }$ \\
SN2005eq & $ 0.09 ^{+ 0.41 }_{- 0.37 }$ & $ -9.67 ^{+ 0.46 }_{- 0.67 }$ & $ -1.44 ^{+ 0.62 }_{- 0.41 }$ \\
SN2005ki & $ 1.10 ^{+ 0.16 }_{- 0.02 }$ & $ -41.55 ^{+ 16.61 }_{- 18.98 }$ & $ 0.06 ^{+ 0.10 }_{- 0.06 }$ \\
SN2006N & $ 1.01 ^{+ 0.08 }_{- 0.13 }$ & $ -24.62 ^{+ 8.57 }_{- 15.50 }$ & $ -0.17 ^{+ 0.17 }_{- 0.26 }$ \\
SN2006S & $ 0.17 ^{+ 0.35 }_{- 0.47 }$ & $ -9.41 ^{+ 0.50 }_{- 0.41 }$ & $ -0.62 ^{+ 0.41 }_{- 0.61 }$ \\
SN2006bt & $ 0.96 ^{+ 0.10 }_{- 0.17 }$ & $ -21.15 ^{+ 6.53 }_{- 13.78 }$ & $ -0.42 ^{+ 0.28 }_{- 0.47 }$ \\
SN2006bz & $ 0.82 ^{+ 0.18 }_{- 0.23 }$ & $ -17.90 ^{+ 4.62 }_{- 11.07 }$ & $ -0.47 ^{+ 0.38 }_{- 0.61 }$ \\
SN2006cm & $ 0.55 ^{+ 0.23 }_{- 0.27 }$ & $ -10.47 ^{+ 0.48 }_{- 0.47 }$ & $ -0.03 ^{+ 0.17 }_{- 0.34 }$ \\
SN2006cq & $ 0.50 ^{+ 0.33 }_{- 0.35 }$ & $ -12.46 ^{+ 2.11 }_{- 6.53 }$ & $ -0.82 ^{+ 0.57 }_{- 0.71 }$ \\
SN2006cs & $ 0.66 ^{+ 0.27 }_{- 0.33 }$ & $ -13.19 ^{+ 2.20 }_{- 7.28 }$ & $ -0.97 ^{+ 0.63 }_{- 0.67 }$ \\
SN2006or & $ 0.25 ^{+ 0.16 }_{- 0.14 }$ & $ -12.47 ^{+ 1.07 }_{- 2.12 }$ & $ 0.05 ^{+ 0.11 }_{- 0.21 }$ \\
SN2007A & $ 0.17 ^{+ 0.32 }_{- 0.37 }$ & $ -9.41 ^{+ 0.41 }_{- 0.35 }$ & $ -0.88 ^{+ 0.38 }_{- 0.47 }$ \\
SN2007N & $ 0.69 ^{+ 0.25 }_{- 0.28 }$ & $ -15.64 ^{+ 3.64 }_{- 9.47 }$ & $ -0.60 ^{+ 0.41 }_{- 0.78 }$ \\
SN2007O & $ 0.34 ^{+ 0.39 }_{- 0.26 }$ & $ -11.90 ^{+ 1.38 }_{- 3.57 }$ & $ -0.60 ^{+ 0.56 }_{- 0.91 }$ \\
SN2007af & $ -0.26 ^{+ 0.41 }_{- 0.31 }$ & $ -9.21 ^{+ 0.35 }_{- 0.37 }$ & $ -0.91 ^{+ 0.38 }_{- 0.45 }$ \\
\hline
\label{sample_1.0_2}
\end{tabular}
\end{table*}

\begin{table*}
\centering
\caption{Host-galaxy properties of our $r=1.0$\,kpc SN sample measured in this work (continued).}
\renewcommand{\arraystretch}{1.5}
%\scriptsize
\begin{tabular}{lrrr}
\hline\hline
SN Name         & $\log \rm Age$ & $\log \rm sSFR$ & $\log \rm Z$ \\
                &    (Gyr)       & ($\rm yr^{-1}$) & ($\rm Z_{\odot}$) \\
\hline
SN2007ba & $ 0.93 ^{+ 0.11 }_{- 0.19 }$ & $ -21.27 ^{+ 6.48 }_{- 13.76 }$ & $ -0.68 ^{+ 0.43 }_{- 0.62 }$ \\
SN2007bz & $ 0.63 ^{+ 0.19 }_{- 0.16 }$ & $ -14.97 ^{+ 2.90 }_{- 6.29 }$ & $ -1.77 ^{+ 0.32 }_{- 0.18 }$ \\
SN2007ci & $ 0.73 ^{+ 0.24 }_{- 0.23 }$ & $ -17.79 ^{+ 4.47 }_{- 10.11 }$ & $ -0.35 ^{+ 0.37 }_{- 0.66 }$ \\
SN2007gi & $ 0.93 ^{+ 0.11 }_{- 0.16 }$ & $ -19.44 ^{+ 4.91 }_{- 12.40 }$ & $ -0.41 ^{+ 0.27 }_{- 0.32 }$ \\
SN2007s & $ 0.29 ^{+ 0.28 }_{- 0.40 }$ & $ -9.61 ^{+ 0.40 }_{- 0.35 }$ & $ -1.35 ^{+ 0.53 }_{- 0.45 }$ \\
SN2008ar & $ 0.54 ^{+ 0.21 }_{- 0.39 }$ & $ -10.20 ^{+ 0.46 }_{- 0.52 }$ & $ -1.51 ^{+ 0.51 }_{- 0.35 }$ \\
SN2008ec & $ -1.33 ^{+ 0.15 }_{- 0.10 }$ & $ -7.87 ^{+ 0.13 }_{- 0.17 }$ & $ 0.19 ^{+ 0.01 }_{- 0.03 }$ \\
SN2009an & $ 0.59 ^{+ 0.27 }_{- 0.24 }$ & $ -15.22 ^{+ 2.61 }_{- 6.31 }$ & $ -0.18 ^{+ 0.25 }_{- 0.64 }$ \\
SN2010ex & $ -0.03 ^{+ 0.31 }_{- 0.17 }$ & $ -10.44 ^{+ 0.87 }_{- 1.01 }$ & $ -0.34 ^{+ 0.42 }_{- 0.85 }$ \\
SN2011ao & $ -0.03 ^{+ 0.41 }_{- 0.38 }$ & $ -9.44 ^{+ 0.35 }_{- 0.38 }$ & $ -1.04 ^{+ 0.36 }_{- 0.45 }$ \\
SN2011hb & $ -0.55 ^{+ 0.37 }_{- 0.41 }$ & $ -8.76 ^{+ 0.52 }_{- 0.46 }$ & $ -1.38 ^{+ 0.61 }_{- 0.46 }$ \\
SN2011ia & $ 0.01 ^{+ 0.46 }_{- 0.33 }$ & $ -9.80 ^{+ 0.33 }_{- 0.49 }$ & $ -1.13 ^{+ 0.44 }_{- 0.49 }$ \\
SN2012cg & $ -0.03 ^{+ 0.11 }_{- 0.10 }$ & $ -10.84 ^{+ 0.25 }_{- 0.25 }$ & $ -0.06 ^{+ 0.18 }_{- 0.35 }$ \\
SN2012da & $ 0.48 ^{+ 0.40 }_{- 0.54 }$ & $ -11.54 ^{+ 2.19 }_{- 9.10 }$ & $ -0.70 ^{+ 0.61 }_{- 0.83 }$ \\
SN2013di & $ 0.36 ^{+ 0.41 }_{- 0.32 }$ & $ -11.91 ^{+ 1.41 }_{- 4.65 }$ & $ -0.72 ^{+ 0.65 }_{- 0.81 }$ \\
\hline
\label{sample_1.0_2}
\end{tabular}
\end{table*}

% If you want to present additional material which would interrupt the flow of the main paper,
% it can be placed in an Appendix which appears after the list of references.

%%%%%%%%%%%%%%%%%%%%%%%%%%%%%%%%%%%%%%%%%%%%%%%%%%

% Don't change these lines
\bsp	% typesetting comment
\label{lastpage}
\end{document}